# Quantum metric induced nonlinear transport in a topological antiferromagnet

Naizhou Wang[1], Daniel Kaplan[2], Zhaowei Zhang[1], Tobias Holder[2], Ning Cao[3], Aifeng Wang[3], Xiaoyuan Zhou[3], Feifei Zhou[1], Zhengzhi Jiang[1], Chusheng Zhang[1], Shihao Ru[1], Hongbing Cai[1,4], Kenji Watanabe[5], Takashi Taniguchi[6], Binghai Yan[2]*, Weibo Gao[1,4,7]*

[1] Division of Physics and Applied Physics, School of Physical and Mathematical Sciences, Nanyang Technological University, Singapore, Singapore

[2] Department of Condensed Matter Physics, Weizmann Institute of Science, Rehovot, Israel

[3] Low Temperature Physics Laboratory, College of Physics and Center for Quantum Materials and Devices, Chongqing University, Chongqing, China

[4] The Photonics Institute and Centre for Disruptive Photonic Technologies, Nanyang Technological University, Singapore.

[5] Research Center for Functional Materials, National Institute for Materials Science, 1-1 Namiki, Tsukuba, Japan

[6] International Center for Materials Nanoarchitectonics, National Institute for Materials Science, 1-1 Namiki, Tsukuba, Japan

[7] Centre for Quantum Technologies, National University of Singapore, Singapore

*Corresponding authors' email: binghai.yan@weizmann.ac.il; wbgao@ntu.edu.sg

**The Berry curvature and quantum metric are the imaginary part and real part, respectively, of the quantum geometric tensor which characterizes the topology of quantum states[1]. The former is known to generate a zoo of important discoveries such as quantum Hall effect and anomalous Hall effect (AHE)[2,3], while the consequences of the quantum metric have rarely been probed by transport. In this work, we observed quantum metric induced nonlinear transport, including both nonlinear AHE and diode-like nonreciprocal longitudinal response, in thin films of a topological antiferromagnet, $MnBi_2Te_4$. Our observation reveals that the transverse and longitudinal nonlinear conductivities reverse signs when reversing the antiferromagnetic order, diminish above the Néel temperature, and are insensitive to disorder scattering, thus verifying their origin in the band structure topology. They also flip signs between electron and hole-doped regions, in agreement with theoretical calculations. Our work provides a pathway to probe the quantum metric through nonlinear transport and to design magnetic nonlinear devices.**

Nonlinear transport provides a powerful tool for probing topological physics in solids[4-21]. A prime example of this is the nonlinear anomalous Hall effect (NLAHE) which reveals the Berry curvature dipole on the Fermi surface and has recently been observed in nonmagnetic topological semimetals (as shown in Fig. 1a)[6,9,10,12,14,15]. This provides a powerful means of investigating quantum geometry in flat space. Another NLAHE caused by the quantum metric, the counterpart of the Berry curvature in the quantum geometry, was also predicted for magnetic topological materials, but is yet to be realized in experiments[1,22-30]. The quantum metric measures the amplitude distance between two neighboring Bloch states and determines the electronic properties of a crystal. A very recent theory[26,30] finds that the quantum metric actually gives rise to both NLAHE and the nonreciprocal longitudinal response, the latter of which is characterized by a diode-like longitudinal resistance, in inversion-breaking magnetic materials (as shown in Fig. 1b)[7]. The asymmetric quantum metric (quantum metric dipole) will induce an anomalous motion of the wave packet and generate nonlinear responses in both the longitudinal and transverse directions. Since the quantum metric is antisymmetric under momentum inversion ($k \rightarrow -k$), both time reversal symmetry ($T$) and inversion symmetry ($P$) must be broken in order to generate a net quantum metric dipole. Additionally, the combined symmetry $PT$ can exclude the contribution from the Berry curvature dipole[4]. As a result, a system which breaks both $P$ and $T$ but preserves $PT$ is expected to be an ideal platform for studying the quantum metric effect (see supplementary information S1 for detailed symmetry analysis).

After careful consideration, we chose $MnBi_2Te_4$ as the candidate platform to investigate the quantum metric. Compared to the magnetic doped topological insulator, the intrinsic magnetic topological insulator, $MnBi_2Te_4$, gained a lot of interest recently due to its unique properties[17,31-41]. As shown in Fig. 1c and e, the crystal structure of $MnBi_2Te_4$ consists of alternating layers of Te-Bi-Te-Mn-Te-Bi-Te, known as septuple layers (SLs)[33,34]. $MnBi_2Te_4$ has an A-type antiferromagnetic (AFM) ground state, in which the Mn spins in each SL are aligned ferromagnetically with an out-of-plane easy axis but are coupled antiparallel to adjacent SLs. For even-layer $MnBi_2Te_4$, both $P$ and $T$ symmetry are broken but the combined $PT$ symmetry is preserved below the Néel temperature[37]. Fig. 1d shows the spatial reflectance magnetic circular dichroism (RMCD) mapping of 3SL and 4SL $MnBi_2Te_4$ measured at zero magnetic field, indicating fully compensated AFM states in even layer $MnBi_2Te_4$[40]. In addition, $MnBi_2Te_4$ possesses three-fold rotational symmetry (from the optical second harmonic generation shown in Fig. 1f), which suppresses the Berry curvature dipole even in the absence of $PT$-symmetry, making the contribution from the quantum metric much easier to observe[30,41].

**Electron nonlinearity from spin order**

To investigate the electron nonlinearity, we fabricated several high-quality, dual gated even-layer $MnBi_2Te_4$ devices. Here, we focus on a 4SL-$MnBi_2Te_4$ device, the schematic structure of which is shown in Fig. 2a. The dual-gate structure allows us to independently control the carrier density and vertical displacement field. As per theoretical predictions, the quantum metric gives rise to both NLAHE and the nonreciprocal longitudinal response[30]. The former is directly measurable via an alternating current (AC) method, while the latter exhibits diode-like nonreciprocal resistance behavior, expressed as: $V_x = R_0(I + \gamma I^2)$ where $I$ represents the applied current, $R_0$ is the linear resistance, and $\gamma$ denotes the coefficient characterizing the nonreciprocity[7]. Considering that the nonreciprocal longitudinal response stems from the quadratic term of the current, we adopted the AC method to measure the second-harmonic voltage $V_x^{2\omega}$ to reflect the non-reciprocal behavior in longitudinal direction, which offers a better signal-to-noise ratio (see supplementary information section S4). To measure the linear and non-linear responses of the device, we employed a standard lock-in technique. As depicted in Fig. 2a, an AC current with frequency of $\omega$ ($I^\omega$) was injected, and the linear voltage $V^\omega$ and second-harmonic voltage $V^{2\omega}$ are simultaneously probed. In our experiment, the $x$-axis is defined as the current direction and the $y$-axis as the transverse direction to the current. All the measurements are conducted at $T = 1.8$ K in the absence of a magnetic field unless otherwise specified. Fig. 2b shows the linear longitudinal ($V_x^\omega$) and transverse ($V_y^\omega$) voltages, indicating good ohmic contact and a negligible misalignment in the Hall bar geometry from the vanishing of $V_y^\omega$.

Next, we focus on the nonlinear responses of 4SL-MnBi$_2$Te$_4$ in opposite AFM states. Utilizing a method akin to those reported previously for controlling AFM states in CrI$_3$ or even-layer MnBi$_2$Te$_4$, we prepared the AFM-I and AFM-II states[37,42] with opposite Néel orders. As shown in Fig. 2c and 2e, the AFM-I state is prepared by sweeping the magnetic field from –7 T to zero, while the AFM-II state is achieved by sweeping the magnetic field from +7 T to zero. For the linear responses, both AFM states exhibited identical longitudinal ($V_x^\omega$) voltages, and the transverse ($V_y^\omega$) voltages remained at zero (Extended Data Fig. 1). This observation aligns with the fully compensated AFM orders in 4SL-MnBi$_2$Te$_4$ (Extended Data Fig. 2). However, the nonlinear responses differ significantly from their linear counterparts. Focusing first on the AFM-I state, Fig. 2c shows that while the linear transverse ($V_y^\omega$) voltage remains at zero, there is a significant negative nonlinear transverse voltage (notated as $V_y^{2\omega}$) that scales quadratically with the injected current $I^\omega$. More importantly, we also detected a prominent negative nonlinear longitudinal voltage (denoted as $V_x^{2\omega}$), with the same order of magnitude as $V_y^{2\omega}$. This observation aligns with the prediction that the quantum metric dipole can induce both nonlinear Hall and nonreciprocal longitudinal response, distinctly different from the response observed in the nonlinear Hall effect caused by the Berry curvature dipole, where only a Hall response is allowed and the longitudinal response is absent[9,10]. The nonreciprocal longitudinal response in MnBi$_2$Te$_4$ can also be observed through DC measurements (see supplementary information section S4). Subsequently, we prepared the AFM-II state. As shown in Fig. 2d, in sharp contrast, although the amplitudes of $V_y^{2\omega}$ and $V_x^{2\omega}$ remain nearly the same with the AFM-I state, the sign of both flips to positive. The sign reversal of the nonlinear response suggests that the nonlinear response observed in MnBi$_2$Te$_4$ is associated with the AFM order. We carefully examined and ruled out potential alternative origins, such as thermal effect or accidental diode junction that could lead to a nonlinear effect. We also exclude the possibility that the nonlinear response observed in even-layer MnBi$_2$Te$_4$ is originated from the residual magnetization (See supplementary information section S2). To further confirm that the nonlinear signal is an intrinsic response originating from the *PT*-symmetric AFM order, we investigate whether the nonlinear response complies with the rotational symmetry of AFM even-layer MnBi$_2$Te$_4$, where $\sigma_{yxx} = -\sigma_{yyy}$ and $\sigma_{xyy} = -\sigma_{xxx}$. When $y$ aligns with the in-plane crystal axis, it should follow that $\sigma_{xyy} = -\sigma_{xxx} = 0$. The coexistence of $\sigma_{yxx}^{2\omega}$ and $\sigma_{xxx}^{2\omega}$ in this 4SL-MnBi$_2$Te$_4$ device suggests some misalignment between *x/y* and the crystal axes, consistent with our optical SHG measurement in Fig. 1f. To substantiate this further, we fabricated another 4SL-MnBi$_2$Te$_4$ device with radially distributed electrodes aligned to the crystalline axis (dashed line in Fig. 2e). Fig. 2f demonstrates the nonlinear transverse and longitudinal response upon rotation of the current injection direction. Both transverse $V_y^{2\omega}$ and longitudinal $V_x^{2\omega}$ exhibit three-fold symmetry. These observed features align consistently with the symmetry of the AFM even-layer MnBi$_2$Te$_4$.

We further investigated the temperature dependence of the nonlinear response in MnBi$_2$Te$_4$. Fig 3a and 3b shows the temperature dependent nonlinear voltages $V_y^{2\omega}$ and $V_x^{2\omega}$ which are prepared at AFM-I and AFM-II states, respectively. As the temperature increases, we found that the nonlinear voltage gradually decreases and vanishes above the Néel temperature of MnBi$_2$Te$_4$, indicating that the nonlinear response is absent in the non-magnetic states. In addition, the nonlinear response in even-layer MnBi$_2$Te$_4$ is only presented at the AFM states and vanishes when all spins are aligned in one direction (see details in supplementary information section S3). In summary, we can conclude that the nonlinear response observed is associated with the nonlinearity of electron motion originating from the AFM order in 4SL-MnBi$_2$Te$_4$.

**Quantum metric induced nonlinear transport**

We now systematically investigate the microscopic origin of the nonlinear response observed in 4SL-MnBi$_2$Te$_4$. To quantify the strength of the nonlinear response, we use the nonlinear conductivity, as it is independent of the sample size. In our experiment, we can extract the longitudinal and transverse nonlinear conductivity from our data as $\sigma_{xxx}^{2\omega} = \frac{J_x^{2\omega}}{(E_{xx}^{\omega})^2} = \frac{V_x^{2\omega}}{(I_x^{\omega})^2 R_{xx}^3} L$ and $\sigma_{yxx}^{2\omega} = \frac{J_y^{2\omega}}{(E_{xx}^{\omega})^2} = \frac{V_y^{2\omega}}{(I_x^{\omega})^2 R_{xx}^3} \frac{L^3}{W^2}$, respectively, where $L$ and $W$ are the longitudinal and transverse length of the Hall bar device. In theory, the nonlinear conductivity due to an applied electric field $E_x$ in a metal includes three contributions[30],

$$J_c = \sigma_{cxx} E_x^2 = (\sigma_{cxx}^2 + \sigma_{cxx}^1 + \sigma_{cxx}^0) E_x^2 \qquad \text{(Eq. 1)}$$

where $\sigma_{cxx}^i$ has the $i$-th order of $\tau$ dependence ($c = x$ or $y$). Here $\sigma_{cxx}^2 = -\frac{e^3 \tau^2}{\hbar^3} \int d^2k f_n \partial_{k_c} \partial_{k_x} \partial_{k_x} \varepsilon_n(\mathbf{k})$ is the nonlinear Drude weight, and $\varepsilon_n(\mathbf{k})$ is the energy of Bloch state $n$ at momentum $\mathbf{k}$, and $f_n$ is the Fermi Dirac distribution for band $n$. $\sigma_{cxx}^1 = -\frac{e^3 \tau}{\hbar^2} \int d^2k f_n (2 \partial_{k_x} \Omega_n^{xc})$, is the Berry curvature dipole contribution[4]. The conductivity due to the normalized dipole of the quantum metric is[30],

$$\sigma_{baa}^0 = -\frac{e^3}{\hbar^2} \int d^2k f_n (2 \partial_{k_b} G_n^{aa} - \partial_{k_a} G_n^{ab}) \qquad \text{(Eq. 2)}$$

Where $a/b = x/y$, $G_n^{ab} = \sum_{m \neq n} \frac{A_{nm}^a A_{mn}^b + A_{nm}^b A_{mn}^a}{(\varepsilon_n - \varepsilon_m)}$ is the band-energy normalized quantum metric and $\partial_{k_a} G_n^{ab}$ represents the normalized quantum metric dipole. In the specific case of 4SL-MnBi$_2$Te$_4$, the *PT*-symmetry excludes any contribution from the Berry curvature dipole[4]. Additionally, the Berry curvature dipole contribution, which is even under time-reversal, contradicts the sign reversal of the nonlinear signal between the AFM-I and II phases. It is also worth noting that the skew scattering mechanism can also contribute to the nonlinear response[21,43], which is significantly suppressed by *PT*-symmetry[44]. Although the anomalous skew scattering is not excluded by *PT*-symmetry[45], we can exclude its contribution based on the scaling analysis discussed subsequently. Therefore, the intrinsic contributions to the nonlinear conductivity in MnBi$_2$Te$_4$ are determined by the nonlinear Drude weight and the normalized quantum metric dipole.

To verify the dominant contribution from the quantum metric, we investigate the scaling relationship between nonlinear conductivities ($\sigma_{yxx}^{2\omega}$ and $\sigma_{xxx}^{2\omega}$) and the linear conductivity $\sigma_{xx}^{\omega}$ (i.e. scattering time $\tau$) as a function of temperature. Fig. 3c shows the temperature dependence of longitudinal conductivity $\sigma_{xx}^{\omega}$ and charge carrier density $n_e$. The charge carrier density $n_e$ remains almost unchanged at the temperature of 1.6 to 10 K, suggesting that the $\tau$ is the main component to determine the conductivity $\sigma_{xx}^{\omega}$ in this range. By adopting this range, we plot the $\sigma_{yxx}^{2\omega}$ and $\sigma_{xxx}^{2\omega}$ as a function of conductivity $\sigma_{xx}^{\omega}$,

$$\sigma^{2\omega} = \eta_2 (\sigma_{xx}^{\omega})^2 + \eta_0 \qquad \text{(Eq. 3)}$$

we find that the predominant contribution to both $\sigma_{xxx}^{2\omega}$ and $\sigma_{yxx}^{2\omega}$ is the intrinsic nonlinear conductivity from the normalized quantum metric dipole ($\eta_0$, $\tau$-independent term). The nonlinear Drude contribution ($\eta_2$, $\tau^2$-dependent term) is nearly negligible compared to the intrinsic one, which is reasonable considering the small conductivity and low mobility in MnBi$_2$Te$_4$. The dominant $\tau$-independent nonlinear conductivity further exclude contribution from the impurity scattering, because its lowest order contribution to the nonlinear conductivity begins at $\tau^1$ (see detailed discussion in supplementary information S7.2) [21,44]. In conclusion, we attribute the nonlinear response observed in 4SL-MnBi$_2$Te$_4$ to the nonlinear Drude weight and the normalized quantum metric dipole, with the latter being the dominant contributor.

**Fermi energy dependent nonlinear response**

We next study how the nonlinear conductivity in 4SL-MnBi$_2$Te$_4$, originating from normalized quantum metric dipole, is influenced by the vertical displacement electric field (denoted as *D*) and the electron charge carrier density (denoted as $n_e$). We first examine the effect of *D* on the nonlinear response. Although *PT* symmetry is instrumental in isolating the quantum metric dipole contribution, the induced nonlinear response remains even when *PT* is broken by *D*. Furthermore, we note that the Berry curvature dipole induced NLAHE still vanishes because of three-fold rotational symmetry. To demonstrate this, we investigate the dependence of $\sigma_{yxx}^{2\omega}$ on the electric field $D$ with fixed carrier density at $n_e \approx -3 \times 10^{12}\ cm^{-2}$. As shown in Fig. 4b, $\sigma_{yxx}^{2\omega}$ changes only slightly when tuning the electric field away from $D=0$. Next, we investigate the effect of charge carrier density $n$ on the nonlinear response. Fig. 4c shows the nonlinear conductivity $\sigma_{yxx}^{2\omega}$ and $\sigma_{xxx}^{2\omega}$ as a function of carrier density $n$, respectively. When the Fermi level is tuned into the charge neutral gap, both $\sigma_{yxx}^{2\omega}$ and $\sigma_{xxx}^{2\omega}$ vanish. This is in line with the fact that the quantum metric dipole is a Fermi surface property[30]. Furthermore, both $\sigma_{yxx}^{2\omega}$ and $\sigma_{xxx}^{2\omega}$ exhibit a sign reversal when tuning the carrier density between hole and electron regimes. Lastly, given the evident nonreciprocal longitudinal response, we can evaluate the nonreciprocity coefficient $\gamma$ in MnBi$_2$Te$_4$. We find the coefficient $\gamma$ can reach $\gamma \approx 7 \times 10^{-11} m^2 A^{-1}$, which is two or three orders larger than traditional heavy metal (HM)/ferromagnetic metal (FM) heterostructures (See details in supplementary information section S4.3) [46,47].

To explain our experimental data, we evaluate $\sigma_{yxx}^{2\omega}$ quantitatively by constructing a slab model for 4SL MnBi$_2$Te$_4$ with density-functional theory (DFT) calculations. Fig. 4d shows the total conductivity as a function of carrier density, in which we assume $\sigma_{yxx}^{2\omega}$ for an ideally aligned device ($y$ aligned with the in-plane crystalline axis). Importantly, around the charge neutrality gap ($\mu = 0$), we observe a sign change which reproduces the experimental findings. (see supplementary information S7.1 for the band-resolved contributions to the quantum metric) The deviation of theoretical $\sigma_{yxx}^{2\omega}$ compared to experimental values may be attributed to the known discrepancy in surface band structure between calculations and experiments on MnBi$_2$Te$_4$. Additionally, in our calculation, we find that the Drude contribution is small within a large energy range around the Fermi level, indicating that the nonlinear conductivity is primarily driven by the quantum metric dipole. We conducted similar measurements on another 6SL-MnBi$_2$Te$_4$ device, as detailed in supplementary information section S6.

As a probe, the quantum metric dipole-driven electron nonlinearity in MnBi$_2$Te$_4$ has several advantages and holds great promise for future applications. First, compared to nonlinearity originating from the Berry curvature dipole or skew scattering, the quantum metric dipole-driven electron nonlinearity is independent of the scattering time, suggesting its robustness against disorder scattering. Second, the quantum metric dipole-driven electron nonlinearity is distinct for opposite AFM spin orders and even robust against small perturbations of external magnetic fields (see details in supplementary information S3), making it a promising candidate for use as a nonlinear magnetic memory device. Finally, the quantum metric dipole-driven nonlinear response is more prominent than the "traditional" nonlinear response from the Berry curvature dipole. We compared the nonlinear response observed in $MnBi_2Te_4$ with other 2D material systems, as presented in Extended Table I. We find that our $MnBi_2Te_4$ devices have a much larger nonlinear conductivity than $WTe_2$ and strained $WSe_2$, only smaller than twisted $WSe_2$ and graphene Moiré superlattices, making $MnBi_2Te_4$ a promising candidate for highly efficient rectifying devices[9,10,13,16,19].

## Methods

### MnBi$_2$Te$_4$ single Crystal growth

Single crystals of MnBi$_2$Te$_4$ were grown using a self-flux method[48,49]. The starting materials used in the single crystal processes are Mn slices, Bi lumps, and Te chunks. MnTe and Bi$_2$Te$_3$ precursors were prepared by reacting the mixed stoichiometric starting materials at 1100 °C and 900 °C for 24 h, respectively. Then MnTe and Bi$_2$Te$_3$ were mixed according to the ratio MnTe: Bi$_2$Te$_3$ = 15: 85, loaded into an alumina crucible, and sealed in a quartz tube. The quartz tube was heated to 650 °C in 10 h, dwelled for 12 h, and slowly cooled to 595 °C at a rate of 1 °C/h to grow the single crystals. Shiny single crystals with a typical size of 3×2×0.5 mm$^3$ can be obtained after centrifuging.

### Device fabrication

The MnBi$_2$Te$_4$ thin flakes were obtained using the Al$_2$O$_3$-assisted exfoliation method[35,50]. First, a 70 nm thick layer of Al$_2$O$_3$ was thermally evaporated onto bulk MnBi$_2$Te$_4$ crystals. Second, the Al$_2$O$_3$ thin film was lifted along with thin flakes of MnBi$_2$Te$_4$, which were cleaved from the bulk crystal using thermal release tape. Then, the stacked MnBi$_2$Te$_4$/Al$_2$O$_3$ was transferred onto a transparent Polydimethylsiloxane (PDMS) film. The number of layers in the MnBi$_2$Te$_4$ thin flakes was determined using optical transmittance measurements. After confirming the layer number, the MnBi$_2$Te$_4$ thin flake samples were transferred to a Si wafer coated with 285 nm of SiO$_2$. To ensure high-quality samples, we used a stencil mask method to deposit metal contacts (Cr/Au) on the samples. A 20-40 nm thick h-BN flake was then transferred to the MnBi$_2$Te$_4$ devices as the top gate dielectric layer, followed by the transfer of a graphite thin flake gate on top of the h-BN/MnBi$_2$Te$_4$ heterostructure. The entire device fabrication process was carried out in a nitrogen-filled glove box with O$_2$ and H$_2$O levels below 1 ppm.

### Nonlinear Electrical transport measurements

The electrical transport measurements were performed in a cryogenic-free cryostat equipped with a superconducting magnet (Cryomagnetic). We used a standard lock-in technique (Zurich MFLI) with excitation frequencies ranging from 10 to 200 Hz to measure both the first and second harmonic signals. The data presented in this manuscript was collected at a low frequency of 17.777 Hz. During the transport measurements, the phase of the first and second harmonic signals was confirmed to be approximately 0 and 90 degrees, respectively (see supplementary information S2.4). The gate voltages were applied using a Keithley 2636 SourceMeter. To independently control the charge carrier density $n_e$ and displacement electric field $D$, we used a dual gated device structure. The charge carrier density $n$ can be obtained by $n = (\frac{\varepsilon_0 \varepsilon^{BN}(V_{TG}-V_{TG0})}{d_{BN}} + \varepsilon_0(V_{BG}-V_{BG0})/(\frac{d_{Al_2O_3}}{\varepsilon^{Al_2O_3}}+\frac{d_{SiO_2}}{\varepsilon^{SiO_2}}))/e$. Here we regarded the Al$_2$O$_3$ and SiO$_2$ dielectric layer as two series connected capacitors. The electric displacement field $D$ can be obtained by $D = (\frac{V_{BG}-V_{BG0}}{(\frac{d_{Al_2O_3}}{\varepsilon^{Al_2O_3}}+\frac{d_{SiO_2}}{\varepsilon^{SiO_2}})} - \frac{\varepsilon^{BN}(V_{TG}-V_{TG0})}{d_{BN}})/2$. Here, $\varepsilon_0$ is the vacuum permittivity; $V_{TG0}$ and $V_{BG0}$ correspond to the gate voltage of maximum resistance; $\varepsilon^{Al_2O_3}$, $\varepsilon^{SiO_2}$ and $\varepsilon^{BN}$ are the relative dielectric constant of Al$_2$O$_3$, SiO$_2$ and h-BN; $d_{Al_2O_3}$, $d_{SiO_2}$ and $d_{BN}$ are the thickness of Al$_2$O$_3$, SiO$_2$ and h-BN, respectively. We noticed a minor asymmetry in $\sigma_{yxx}^{2\omega}$ and $R_{xx}$ with respect to the displacement electric field $D$, possibly attributed to charge trapping induced by substrate effects or disorder[51]. The asymmetry is more pronounced in $\sigma_{yxx}^{2\omega}$, as the quantum metric is related to energy differences between bands and more sensitive to $D$.

**The nonlinear response in the MnBi$_2$Te$_4$ device with radially distributed devices**

For the MnBi$_2$Te$_4$ device with radially distributed devices, when the current aligns with the in-plane crystal axis ($\theta = 0°$), the transverse $V_y^{2\omega}$ is zero. In contrast, when the current is perpendicular to the crystalline axis ($\theta = 90°$), the longitudinal $V_x^{2\omega}$ is zero. In addition, the nonlinear response should obey the symmetry of even-layer MnBi$_2$Te$_4$, with $\sigma_{yxx} = -\sigma_{yyy}$ and $\sigma_{xyy} = -\sigma_{xxx}$. To substantiate this, we use the formula for nonlinear conductivity: $\sigma_{xxx}^{2\omega} = \frac{J_x^{2\omega}}{(E_{xx}^{\omega})^2} = \frac{V_x^{2\omega} L}{(I_x^{\omega})^2 R_{xx}^3}$ and $\sigma_{yxx}^{2\omega} = \frac{J_y^{2\omega}}{(E_{xx}^{\omega})^2} = \frac{V_y^{2\omega}}{(I_x^{\omega})^2 R_{xx}^3}\frac{L^3}{W^2}$. Here, $L$ refers to the length and $W$ to the width of the device. The optical image (Fig. 2e) reveals that the length-to-width ratio, $L/W$, is approximately 0.85. Taking into consideration the geometric characteristics of the device, we conclude that the nonlinear conductivity conforms to $\sigma_{yxx} = -\sigma_{yyy}$ and $\sigma_{xyy} = -\sigma_{xxx}$.

**Optical measurement**

For the RMCD measurement, a linearly polarized 633 nm HeNe laser was modulated by a photoelastic modulators (PEM) at a frequency of 50 kHz. The laser beam was focused on the sample at normal incidence using an Olympus MPLN50× objective with a 0.75 numerical aperture. The laser spot size was approximately 2 µm, with a power of 1.8 µW. The reflected light was collected by a photodetector and analyzed using a lock-in amplifier set to the same frequency as the PEM. The RMCD measures the differential reflection between left and right circularly polarized light, the magnitude of which is proportional to the magnetic moments of the MnBi$_2$Te$_4$ sample.

The second harmonic generation (SHG) measurement was performed at room temperature using femtosecond pulse lasers with a central wavelength of 800 nm. The laser beam was focused at normal incidence on the sample using a Nikon ELWD 100x microscope objective. The SHG signals were collected using a spectrometer. To perform the polarization-resolved measurement, a motorized polarizer was used to control the polarization of the incident laser beam as it was rotated next to the objective.

**DFT calculations**

We use VASP (Vienna ab-initio software package) with the PBE functional in the generalized gradient approximation, to obtain the electronic ground state[52]. We then project the ground state wavefunctions on Wannier functions using Wannier 90[53] and create a tight-binding model with 184 Wannier orbitals. For the Drude contribution we adopt a lifetime of $\tau \approx 40\ fs$ estimated from our experimental conductivity and carrier density in Fig. 2c and theoretical effective mass ($m^* = 0.14 m_e$) for the 4 SL film.

**Acknowledgements:** We thank Shengyuan A. Yang, Huiying Liu, Suyang Xu and Justin Song for helpful discussion. W.-b.G acknowledges the financial support from the Singapore National Research Foundation through its Competitive Research Program (CRP Award No. NRF-CRP22-2019-0004). B.Y. acknowledges the financial support by the European Research Council (ERC Consolidator Grant No. 815869, "NonlinearTopo") and Israel Science Foundation (ISF No. 2932/21). A. F. Wang acknowledges the financial support from the National Natural Science Foundation of China (Grant No. 12004056), Chongqing Research Program of Basic Research and Frontier Technology, China (Grants No. cstc2021jcyj-msxmX0661) and Fundamental Research Funds for the Central Universities, China (Grant No. 2022CDJXY-002). X.Z acknowledges the financial support from the National Key Research and Development Program of the Ministry of Science and Technology of China (2019YFA0704901) and the National Natural Science Foundation of China (Grant Nos. 52125103, 52071041).

**Author contributions:** W.-b.G and B.Y conceived and supervised the project. N.W fabricated the devices, performed the transport and RMCD measurement with help from Z.Z and C.Z. D.K, T.H and B.Y performed the theoretical calculations. N.W and F.Z performed the NV measurement with help from Z.J, S.R and H.C. N.W, W.-b.G, D.K, T.H and B.Y analysed the data. N.C, A.W and X.Z grew the MnBi$_2$Te$_4$ single crystals. K.W and T.T grew the hBN single crystals. N.W, B.Y and W.-b.G wrote the manuscript with input from all authors.

**Competing interests**: The authors declare no competing interests.

**Additional information:** Correspondence and requests for material should be addressed to Binghai Yan (binghai.yan@weizmann.ac.il) or Wei-bo Gao (wbgao@ntu.edu.sg).

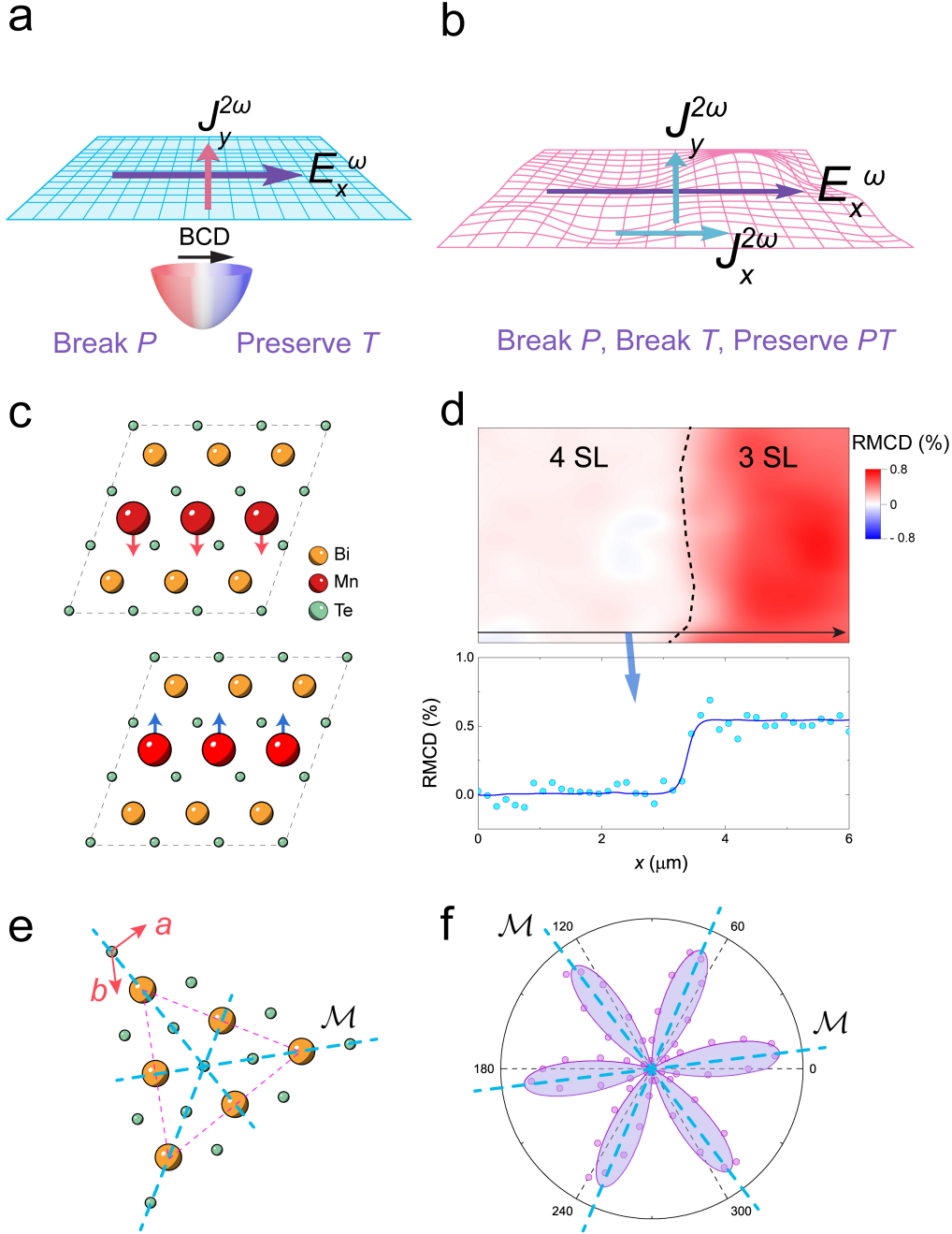

**Figure 1 Quantum metric induced nonlinearity in a *PT*-symmetric AFM. a.** The nonlinear Hall effect originating from Berry curvature dipole, which has been observed in systems with broken inversion symmetry. **b.** The nonlinear longitudinal and transverse response observed in a *PT*-symmetric system due to quantum metric dipole. **c.** The crystal structure of even-layer MnBi$_2$Te$_4$ from the side view. The arrows represent the spin momentum direction of the Mn atoms. **d.** The spatial reflectance magnetic circular dichroism (RMCD) mapping of MnBi$_2$Te$_4$ measured at zero magnetic field. The dashed line marks the boundary between the 4SL-MnBi$_2$Te$_4$ and 3SL-MnBi$_2$Te$_4$. **e.** The crystal structure of MnBi$_2$Te$_4$ from the top view. The dashed line marked with *M* represents the mirror line. **f.** The angle dependent optical second-harmonic generation (SHG) of the 4SL-MnBi$_2$Te$_4$ sample. The zero degree corresponds to the applied current direction in the device.

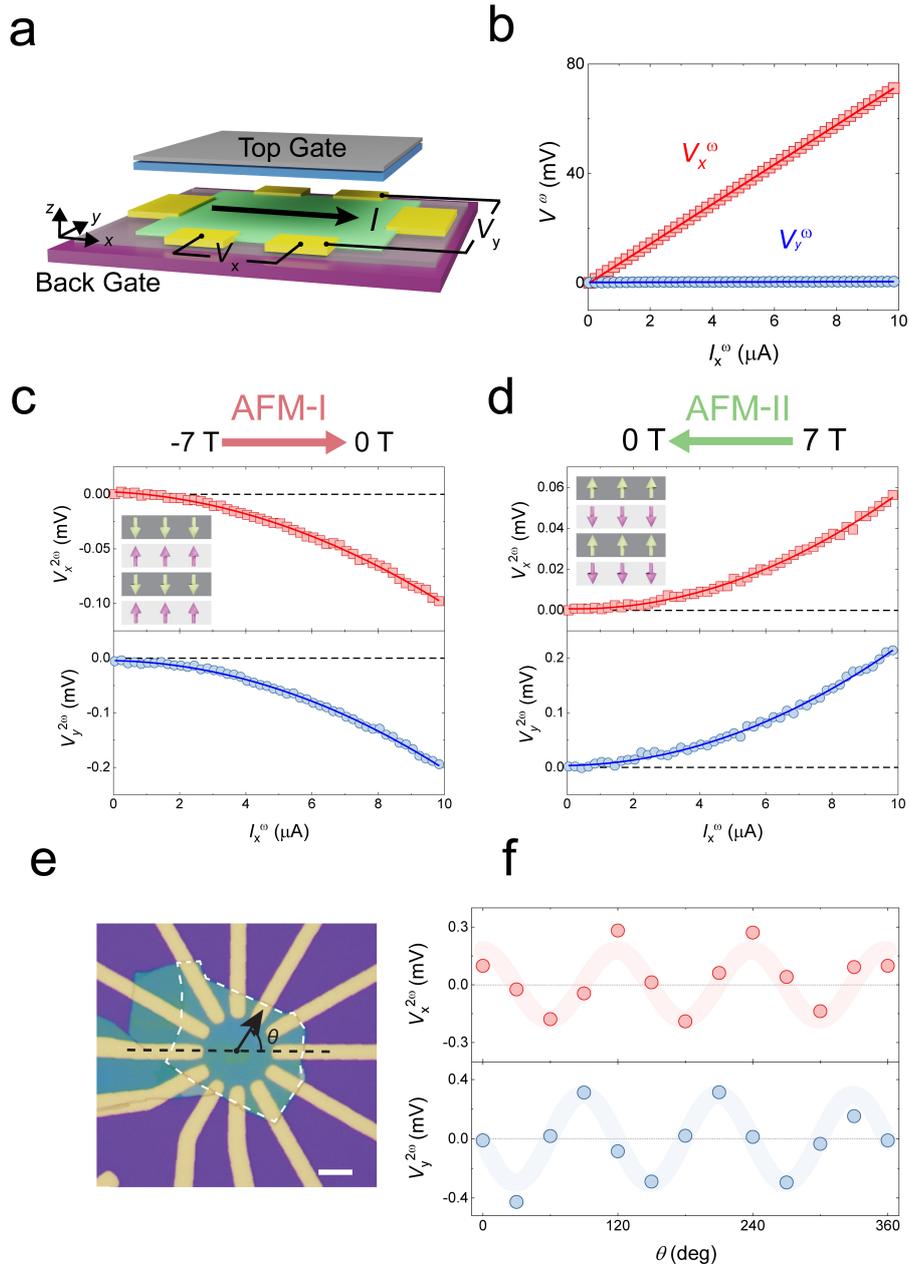

**Figure 2 Observation of nonlinear transport in AFM MnBi$_2$Te$_4$. a.** The schematic view of the dual-gated 4SL-layer MnBi$_2$Te$_4$ device. The linear and nonlinear signals are measured simultaneously with the denoted direction. **b.** The linear longitudinal $V_x^\omega$ and transverse $V_y^\omega$ voltage as a function of current $I_x^\omega$. The solid line is the linear fitting of the data. **c and d.** The nonlinear longitudinal $V_x^{2\omega}$ and transverse $V_y^{2\omega}$ voltage as a function of current $V_x^\omega$ for AFM-I and AFM-II states, respectively. At a fixed electric field, the different AFM states are prepared by: AFM-I, sweeping the magnetic field from -7 T to 0 T; AFM-II, sweeping the magnetic field from +7 T to 0 T. The solid lines are quadratic fits to the data. **e.** The optical image of another 4SL-MnBi$_2$Te$_4$ device with radially distributed electrodes. The dashed line indicates the in-plane crystalline axis, determined by SHG measurement. scale bar: 5 $\mu$m. **f.** The nonlinear transverse $V_y^{2\omega}$ and longitudinal $V_x^{2\omega}$ response as a function of current injection angle $\theta$.

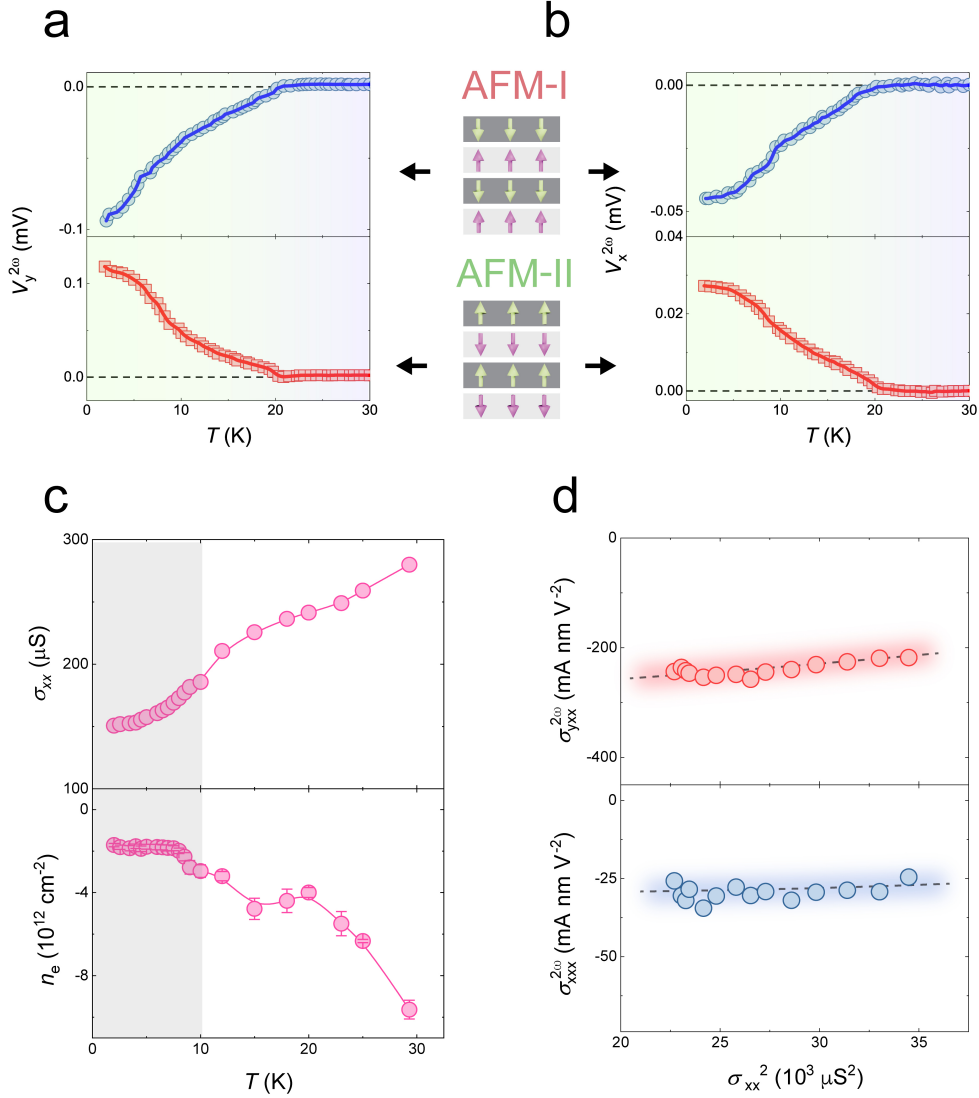

**Figure 3 Spin order related electron nonlinearity from band-normalized quantum metric dipole. a.** Nonlinear transverse voltage $V_y^{2\omega}$ of 4SL-MnBi$_2$Te$_4$ as a function of temperature for opposite AFM-I and AFM-II states, respectively. The amplitudes of $V_y^{2\omega}$ are similar for the two AFM states but their sign is reversed. $V_y^{2\omega}$ vanishes when the temperature is above the Néel temperature of MnBi$_2$Te$_4$. **b.** The nonlinear longitudinal voltage $V_x^{2\omega}$ as a function of temperature for opposite AFM-I and AFM-II states, respectively, exhibiting similar trend as $V_y^{2\omega}$. **c.** The conductivity and fitted carrier density of 4SL-MnBi$_2$Te$_4$ at different temperatures. **d.** The scaling relationship between the nonlinear conductivity $\sigma_{yxx}^{2\omega}$ ($\sigma_{xxx}^{2\omega}$) and the square of the linear longitudinal conductivity $\sigma_{xx}^2$. The scaling is carried out at the temperature range of 2 -10 K (marked with shadow area in **c**), in which the carrier density nearly remains constant. The nonlinear transverse and longitudinal conductivities are extracted by $\sigma_{yxx}^{2\omega} = \frac{V_y^{2\omega}}{(I_x^\omega)^2 R_{xx}^3} \frac{L^3}{W^2}$ and $\sigma_{xxx}^{2\omega} = \frac{V_x^{2\omega} L}{(I_x^\omega)^2 R_{xx}^3}$ , respectively. The dashed line is a fit of the data with the scaling relation $\sigma^{2\omega} = \eta_2 (\sigma_{xx}^\omega)^2 + \eta_0$. From the fitting, we conclude that the predominant contribution is from the $\eta_0$ term, namely from the intrinsic nonlinear conductivity due to the band-normalized quantum metric dipole.

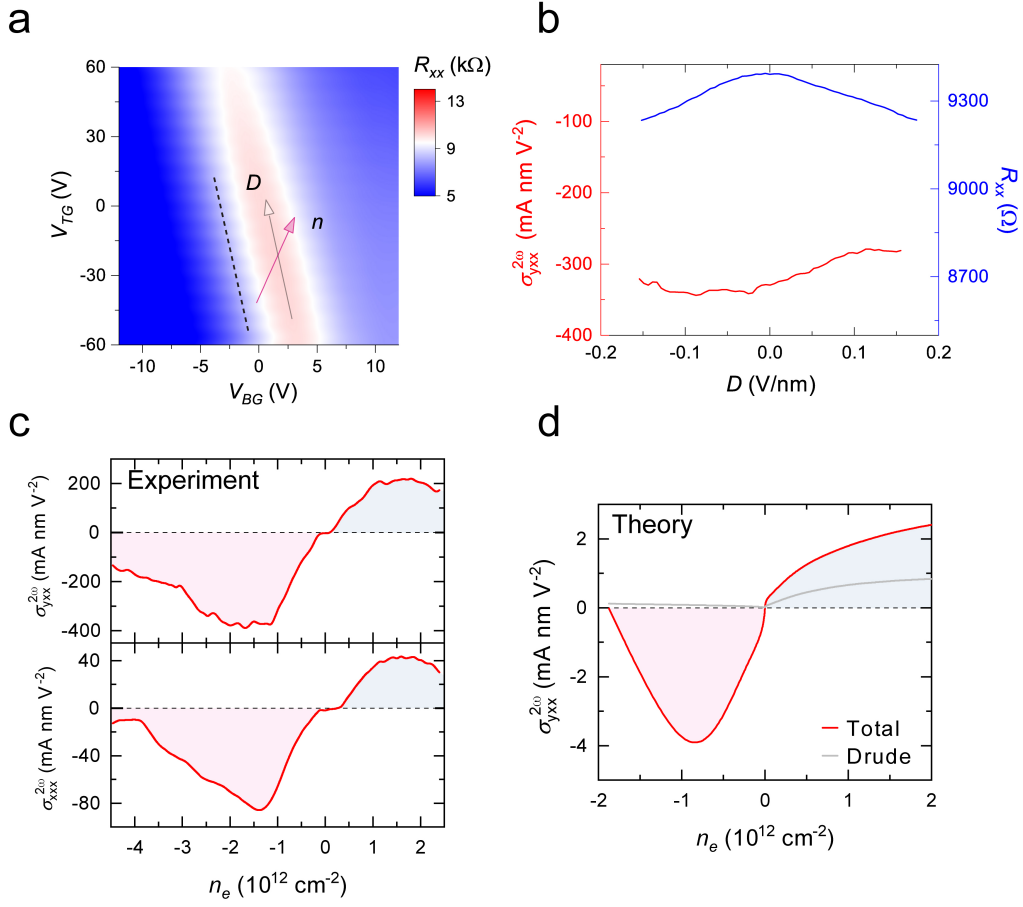

**Figure 4 The electric field and carrier density dependence of the nonlinear response. a.** The resistance map of the 4SL-MnBi$_2$Te$_4$ device. The vertical displacement field $D$ and charge carrier density $n_e$ can be independently controlled, as denoted in the figure. **b.** The nonlinear transverse conductivity $\sigma_{yxx}^{2\omega}$ and longitudinal resistance $R_{xx}$ as a function of vertical displacement electric field. The carrier density is fixed at $n_e \approx -3 \times 10^{12}\ cm^{-2}$ as $D$ varies along the dashed line in **a**. **c.** The measured nonlinear transverse conductivity $\sigma_{yxx}^{2\omega}$ and longitudinal conductivity $\sigma_{xxx}^{2\omega}$ as a function of charge carrier density $n_e$. **d.** The calculated total nonlinear transverse conductivity $\sigma_{yxx}^{2\omega}$ (red line) as a function of carrier density. The grey line denotes the Drude contribution.

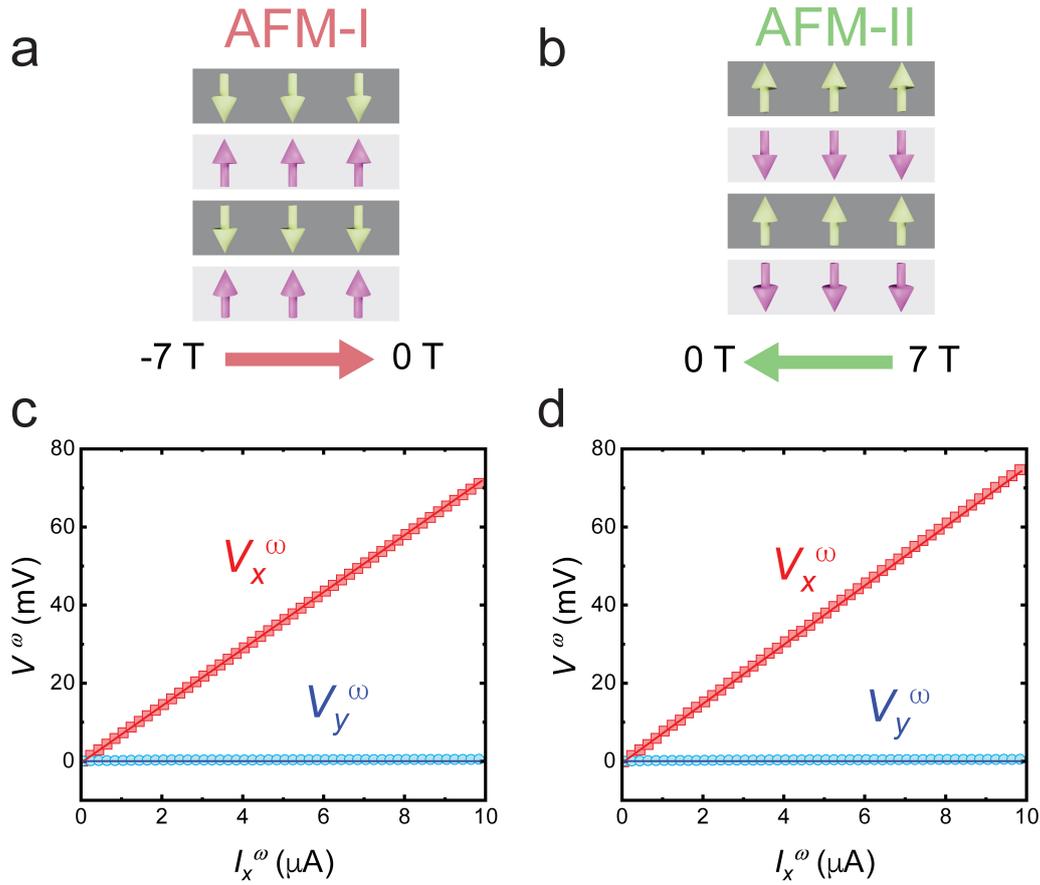

**Extended Figure 1 The linear conductivity of 4SL-MnBi$_2$Te$_4$ with opposite AFM states. a and b.** The AFM-I and AFM-II states are prepared by sweeping the magnetic field from -7 T to 0T or +7 T to 0 T, respectively. **c and d.** The linear longitudinal $V_x^\omega$ and transverse $V_y^\omega$ voltage as a function of current $I_x^\omega$ for AFM I and AFM II states, respectively. The solid line is a linear fit to the experimental data.

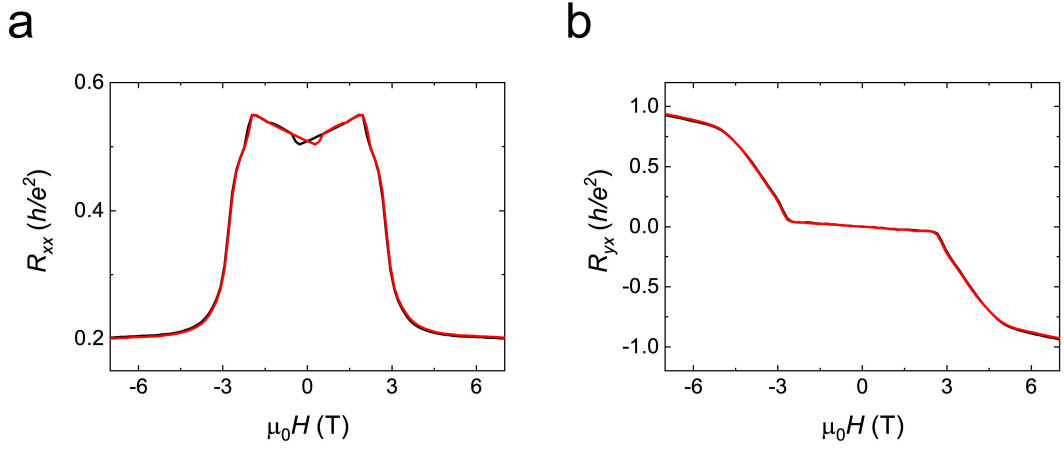

**Extended Figure 2 The fully compensated AFM order in 4SL-MnBi$_2$Te$_4$ device. A.** The magnetic field dependent longitudinal resistance $R_{xx}$ of the 4SL-MnBi$_2$Te$_4$ device. **B.** The magnetic field dependent Hall resistance $R_{yx}$ of the 4SL-MnBi$_2$Te$_4$ device. In zero magnetic field, the AFM order is fully compensated and the Hall resistance $R_{yx} = 0$.

| Material | $T$ (K) | $\sigma_{xx}^{1\omega}$ ($10^{-3}$S) | $\sigma_{yxx}^{2\omega}$ (mA nm V$^{-2}$) | Origin |
|---|---|---|---|---|
| WTe$_2$ (bilayer)[7] | 10 | 0.15 | 940 | BCD |
| WTe$_2$ (few-layer)[6] | 2 | 2.1 | 3 | BCD |
| Strained WSe$_2$[10] | 140 | ~ 0.41 | ~ 1200 | BCD |
| Twisted WSe$_2$[16] | 1.5 | ~ 0.3 | ~ 18000 | BCD |
| h-BN/graphene/h-BN[13] | 1.7 | 17.8 | $1.05 \times 10^7$ | Skew scattering |
| MnBi$_2$Te$_4$ (4-SL) | 2 | 0.15 | 400 | Quantum metric dipole |
| MnBi$_2$Te$_4$ (6-SL) | 2 | 0.5 | 2400 | Quantum metric dipole |

**Extended Table 1 Comparison of the nonlinear conductivity for MnBi$_2$Te$_4$ and other two-dimensional material systems.** The nonlinear transverse conductivity $\sigma_{yxx}^{2\omega}$ in even-layer MnBi$_2$Te$_4$ is summarized and compared with other reported material systems[9,10,13,16,19]. The corresponding mechanisms responsible for the nonlinear response are also listed for comparison.

**Supplementary information for**

# Quantum metric induced nonlinear transport in a topological antiferromagnet

S1.  Symmetry analysis of the nonlinear response induced by quantum metric dipole

S2.  Analysis of other extrinsic origins for nonlinear response

S3.  Relationship between the magnetic states and the nonlinear response in even-layer MnBi$_2$Te$_4$

S4.  The nonreciprocal longitudinal response in 4SL-MnBi$_2$Te$_4$

S5.  Scaling relationship between the $|\frac{\sigma_{xx}}{n_e}|$ and temperature in 4SL-MnBi$_2$Te$_4$

S6.  Additional data from a 6SL-MnBi$_2$Te$_4$ sample

S7.  The influence of disorder and other mechanism on the nonlinear response in MnBi$_2$Te$_4$

# S1. Symmetry analysis of the nonlinear response induced by quantum metric dipole

In this section, we present the symmetry analysis of the nonlinear response induced by quantum metric dipole. Table S1 provides a comprehensive symmetry analysis of the nonlinear conductivities, paying particular emphasis on the difference between the quantum metric dipole (QMD) and Berry curvature dipole (BCD). In a 2D system, which is directly relevant to our experiment, the longitudinal ($\sigma_{xxx}$) and transverse (Hall) ($\sigma_{yxx}$) nonlinear responses are summarized in Table S1 for several key symmetries. These symmetries act on both lattice and spin.

|  | $P$ (inversion) | $T$ (time reversal) | $PT$ | $C_3$ | $M_x$ ($x$ to -$x$) |
|---|---|---|---|---|---|
| BCD | All forbidden | Allowed, only Hall. | All forbidden | All forbidden | $\sigma_{yxx}$ allowed $\sigma_{xyy} = 0$ |
| QMD | All forbidden | All forbidden | All allowed | Allowed $\sigma_{xxx} = -\sigma_{xyy}$ $\sigma_{yyy} = -\sigma_{yxx}$ | $\sigma_{yxx}$ allowed $\sigma_{xyy} = 0$ |

**Table S1. Symmetry analysis of the nonlinear response originated from quantum metric dipole and Berry curvature dipole**

We elaborate more about general AFM materials. The minimal requirement for the quantum metric dipole-induced nonlinear response is to break both *P* and *T*. This is fulfilled in even layers of AFM MnBi$_2$Te$_4$. However, this response vanishes in bulk MnBi$_2$Te$_4$ due to the presence of inversion symmetry, *P*. Many AFM materials, especially in the 3D bulk phase, have inversion symmetry and cannot show such nonlinear effects. For a general material, one can check its magnetic point groups and verify whether inversion symmetry appear. Another important symmetry is *T*, time-reversal symmetry. AFM or FM breaks *T*, given that *T* reversed all spins by 180 degrees. Consequently, the quantum metric dipole-induced nonlinear response will reverse sign if it exists (with inversion-breaking).

**S2. Analysis of other extrinsic origins for nonlinear response**

In this section, we thoroughly examine potential extrinsic factors that could contribute to the nonlinear response in even-layer MnBi$_2$Te$_4$. After careful analysis, we have determined that these factors can be excluded from the discussion.

❖ **S2.1 Thermoelectric effect from Joule heating**

When a current passes through a sample with external asymmetry (e.g. asymmetric flake shape, contact resistance, asymmetry in external circuit), joule heating can create a temperature gradient ($\Delta T$) across the sample. This temperature gradient can generate a thermoelectric voltage ($V_{thermal}$) described by $V_{thermal} \propto \Delta T \propto I^2 R$. This means that the thermoelectric effect is a second order nonlinear effect, and the thermoelectric voltage is proportional to the resistance of the sample. However, our $V^{2\omega}$ data vanishes above the Néel temperature of MnBi$_2$Te$_4$, while the resistance only shows a small jump at this temperature.

Additionally, the thermoelectric effect caused by the temperature gradient $\Delta T$ can generally be divided into two categories: The Seebeck effect, where $V_{thermal}$ is parallel to the $\Delta T$ direction; The Nernst effect, where $V_{thermal}$ is transverse to the $\Delta T$ direction. The Seebeck effect is even under time reversal ($\mu_0 H \rightarrow -\mu_0 H$), which contradicts our observation of sign reversal when preparing different AFM states. The Nernst effect is odd under time reversal, but it vanishes in fully compensated antiferromagnets with zero net magnetization as in the case for even-layer MnBi$_2$Te$_4$.

Finally, the three-fold rotational symmetry of the nonlinear response $V^{2\omega}$, as illustrated in Fig. 2 e and f, exclude the Joule heating effect to the nonlinear response. Since the joule heat effect must couple with external symmetry breaking, it should never show the three-fold rotational symmetry.

Therefore, we can exclude the thermoelectric effect in our experiment based on the following reasons: (1) Our $V^{2\omega}$ data vanishes above the Neél temperature of MnBi$_2$Te$_4$. (2) Our $V^{2\omega}$ data shows sign reversal when preparing different AFM states. (3) The three-fold rotational symmetry of the nonlinear response $V^{2\omega}$.

❖ **S2.2 Accidental contact junctions**

If a contact diode exists between the electrode and sample, it will produce a rectification effect. However, the extrinsic signals from this source should be strongly related to the contact electrodes and can be tested using different combinations of contact electrodes.

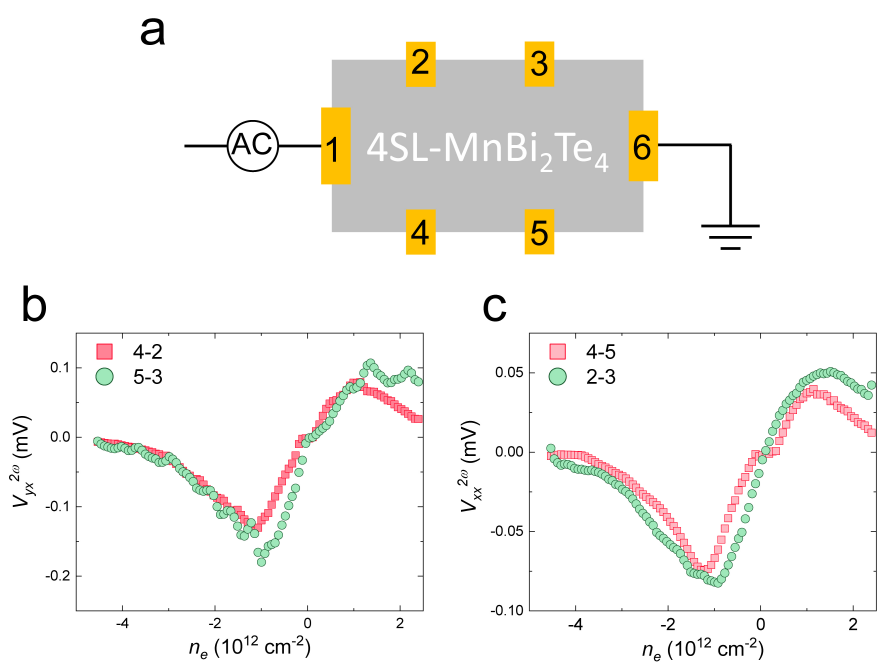

**Fig. S1. The nonlinear voltage of different contacts as a function of carrier density. a.** The schematic device structure of our 4SL-MnBi$_2$Te$_4$ device. The current is injected between the electrodes 1 and 6. **b.** The transverse nonlinear voltage for different electrodes as a function of carrier density. **c.** The longitudinal nonlinear voltage for different electrodes as a function of carrier density.

Figure S1 shows the carrier density dependence of the nonlinear voltage measured with different contacts. With different combinations of transverse electrodes or longitudinal electrodes, the dependence of nonlinear voltage on carrier density exhibits similar behavior and nearly the same amplitudes, indicating no dependence on the contacts. In addition, the extrinsic signals from this source should not show any dependence on the different AFM states. Additionally, we carried out the two-wire DC measurement of the 4SL-MnBi$_2$Te$_4$ sample, as shown in Fig. S2. The observed linear *I-V* characteristics suggest ohmic contact in our device, thereby excluding the extrinsic diode effect.

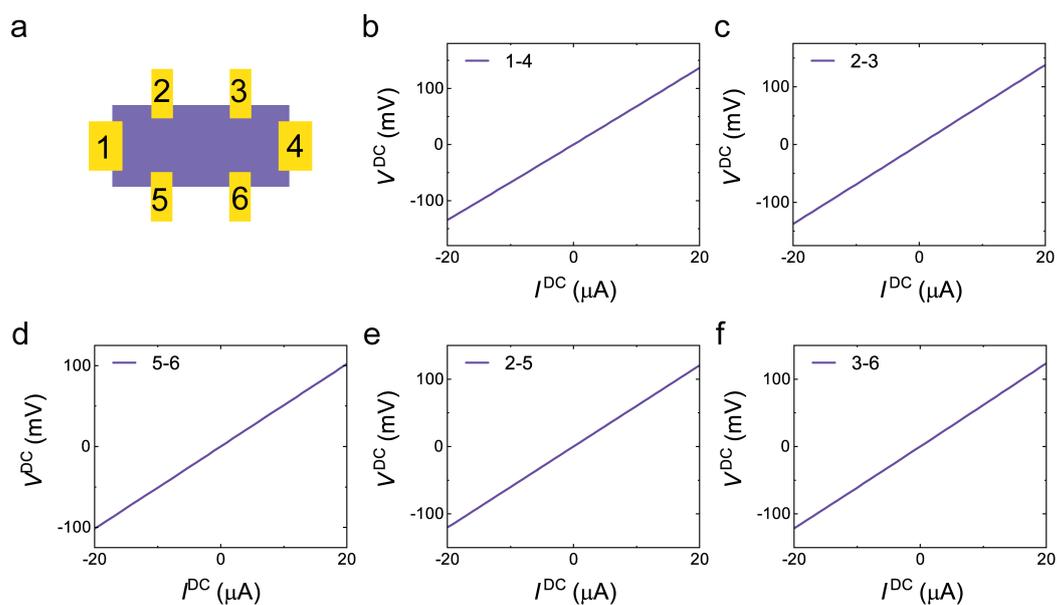

**Fig. S2. Two-terminal DC measurement of the 4SL-MnBi$_2$Te$_4$ device a.** Schematic illustration of the 4SL-MnBi$_2$Te$_4$ device with labeled electrode numbers. **b to f.** Two-terminal DC voltage plotted against current for various electrode configurations.

Therefore, we can exclude the accidental contact junction effect in our experiment based on the following reasons: (1) Our $V^{2\omega}$ data do not show obvious contact dependence. (2) Our $V^{2\omega}$ data show a clear dependence on the different AFM states. (3) Two-wire DC measurements show linear *I-V* characteristics, suggesting ohmic contact and excluding extrinsic diode effect.

## S2.3 Frequency-dependence of the nonlinear response

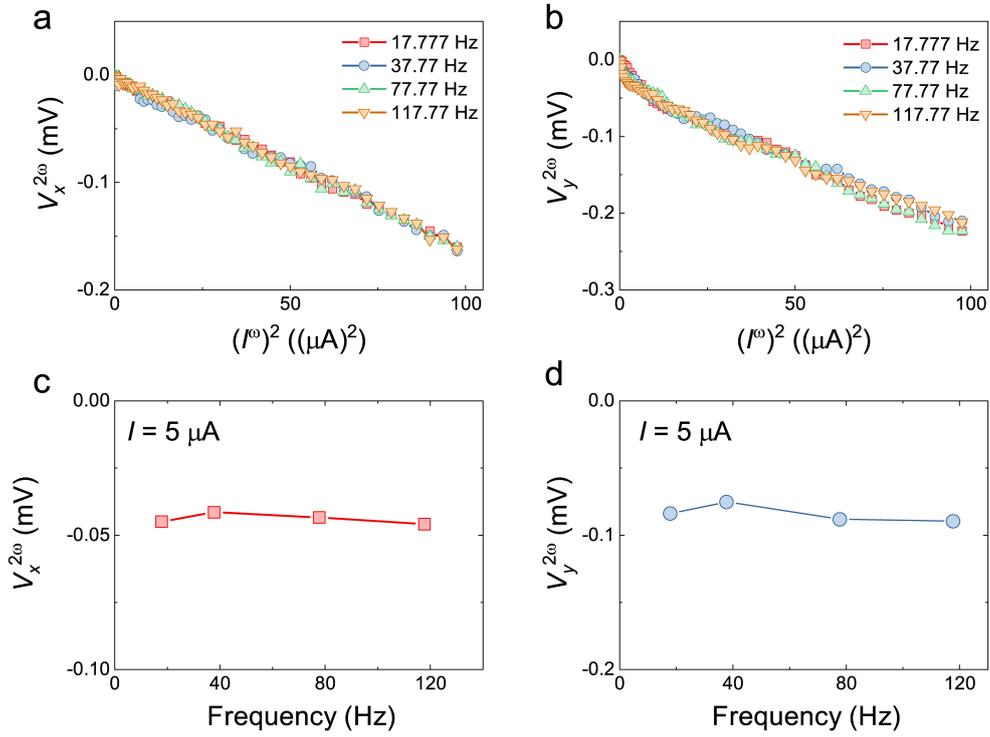

**Fig. S3 Nonlinear Responses of 4SL-MnBi$_2$Te$_4$ at Different Frequencies a.** The nonlinear longitudinal voltage as a quadratic function of applied current across varying frequencies. **b.** The nonlinear transverse voltage as a quadratic function of applied current across varying frequencies. **c.** The nonlinear longitudinal voltage at different frequencies with a fixed current of $I$ = 5 μA. d. The nonlinear transverse voltage at different frequencies with a fixed current of $I$ = 5 μA.

Fig. S3 displays the nonlinear response of 4SL-MnBi$_2$Te$_4$ under different frequencies. The observed nonlinear longitudinal and transverse voltages do not show a frequency dependency (17.777Hz ~ 117.77Hz), excluding the spurious capacitive coupling effect.

## ❖ S2.4. Phase information during the lock-in measurement

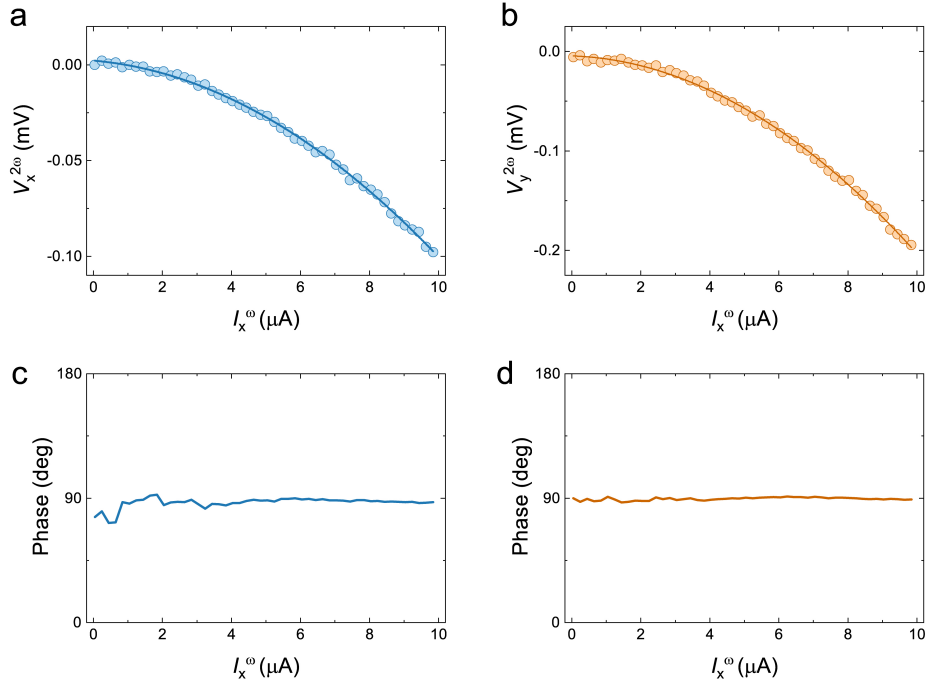

**Fig. S4 Phase information in lock-in measurements a.** The nonlinear longitudinal voltage as a function of the applied current. **b.** The nonlinear transverse voltage as a function of the applied current. **c.** Phase information obtained during measurement of nonlinear longitudinal voltage. **d.** Phase information obtained during measurement of nonlinear transverse voltage.

Fig. S4 presents the phase information during the nonlinear measurement of the 4SL-MnBi$_2$Te$_4$. The lock-in amplifier phase is locked at ~ 90 degrees during the second-order nonlinear measurement, consistent with the expected value for the second-order response.

## ❖ S2.5. Uncompensated magnetic moment in the even-layer MnBi$_2$Te$_4$

To exclude the possibility that the nonlinear responses in even-layer MnBi$_2$Te$_4$ are originated from uncompensated magnetic moment, in this section, we directly measure the magnetism of MnBi$_2$Te$_4$ thin flakes using the nitrogen vacancy (NV) center magnetometry. Moreover, we measure the nonlinear responses of the odd-layer MnBi$_2$Te$_4$ sample, where the nonlinear Hall response is vanished.

- **S2.5.1. Nitrogen vacancy (NV) center magnetometry measurements of the MnBi$_2$Te$_4$ thin flakes**

Here, we utilized a nitrogen vacancy (NV) ensemble in diamond as a sensitive magnetometer to directly assess the magnetic properties of MnBi$_2$Te$_4$. The NV-based magnetometer, known for its impressive sensitivity and excellent spatial resolution, has been recognized as a promising quantum sensor[1]. It has been notably effective in investigating magnetism in 2D materials[2-4]. The ground-state energy level structure of the NV center is illustrated in Fig. S5a, which consists of triplet states $m_s = 0$ ($|0\rangle$) and $m_s = \pm 1$ ($|\pm 1\rangle$) [1]. When an external magnetic field $B_{NV}$ (akin to the net magnetization in ferromagnetic materials) is applied, the $m_s = \pm 1$ ground states ($|\pm 1\rangle$) of the NV center undergo a Zeeman splitting with energy of $2\gamma_{NV}B_{NV}$. Here, $\gamma_{NV} \approx 28.03$ MHz/mT denotes the gyromagnetic ratio of the NV center. By applying the resonant microwave field, the spin transitions $|0\rangle \leftrightarrow |-1\rangle$ and $|0\rangle \leftrightarrow |1\rangle$ can be coherently realized. The corresponding resonances can be determined by sweeping the MW frequency and measuring the spin-state-dependent fluorescence of NV center, as depicted in Fig. S5b. This method is known as optically detected magnetic resonance (ODMR) technique. Therefore, by comparing the differences between $v_1$ and $v_2$, we can extract the magnetic field $B_{NV}$ and determine the magnetic properties of MnBi$_2$Te$_4$.

The NV measurement of MnBi$_2$Te$_4$ is carried out in a 4K optical cryostation. In our experiment specifically, a tiny bias magnetic field $B_{NV}^{bias}$ (~ 8.9 Gauss) is applied to separate the $|\pm 1\rangle$ states for a successful double-Lorentzian fitting. The magnetism in MnBi$_2$Te$_4$ can be described as $[(v_2-v_1)/(2\gamma_{NV})-B_{NV}^{bias}]/\cos\theta$, where $\theta \approx 54.7°$ denotes the angle between the NV axis and the magnetization direction of MnBi$_2$Te$_4$. We conducted ODMR spectroscopy on different single layers (SLs) of the MnBi$_2$Te$_4$ sample (Fig. S5c) and summarized the calculated magnetic field $B_{NV}$ in Fig. S5d. No external magnetic field was detected in even-layer MnBi$_2$Te$_4$. However, an external magnetic field of approximately 30 μT was observed in the 3SL and 5SL MnBi$_2$Te$_4$ samples, thereby ruling out the possibility of residual magnetization in even-layer MnBi$_2$Te$_4$ samples.

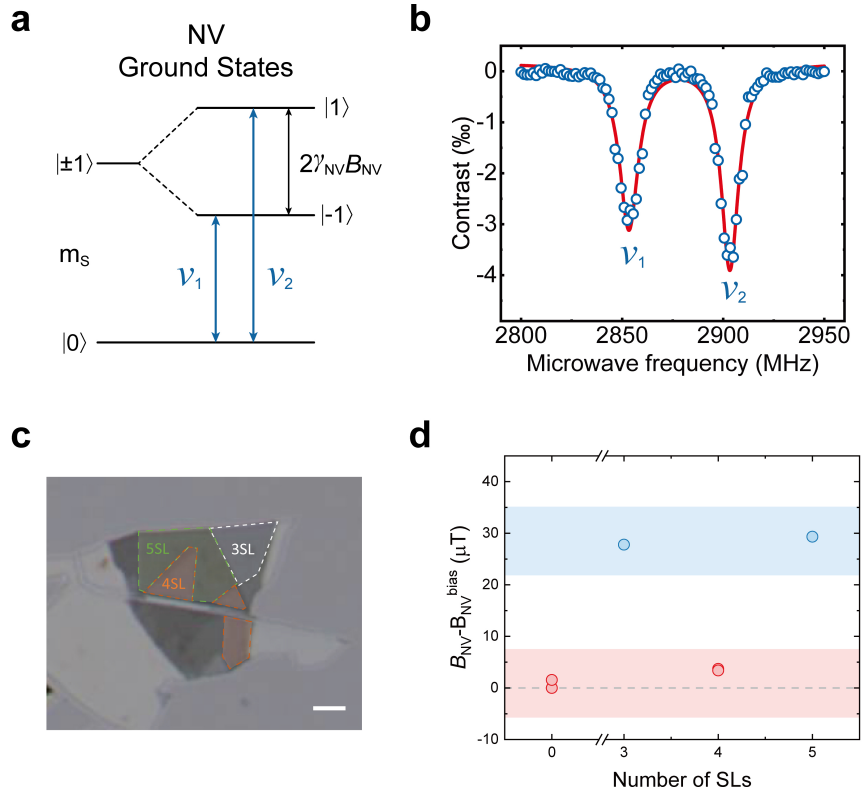

**Fig. S5 Nitrogen vacancy (NV) center magnetometry measurements of the MnBi$_2$Te$_4$ thin flakes a.** The ground-state energy level structure of the NV center, consisting of triplet states $m_s = 0$ ($|0\rangle$) and $m_s = \pm 1$ ($|\pm 1\rangle$). In the presence of external magnetic field $B_{NV}$, like the net magnetization in ferromagnetic materials, the $|\pm 1\rangle$ ground states of NV will experience a Zeeman splitting with energy of $2\gamma_{NV}B_{NV}$. $v_1$ and $v_2$ corresponding to the spin transitions $|0\rangle \leftrightarrow |-1\rangle$ and $|0\rangle \leftrightarrow |1\rangle$. By simultaneously applying 532-nm laser and sweeping microwave frequency, one can observe resonance in fluorescence intensity with the energy (frequency) difference between the $|-1\rangle$ and $|1\rangle$ states. **b.** The ODMR spectrum of the diamond surface. A tiny bias magnetic field $B_{NV}^{bias}$ (~ 8.9 Gauss) is applied to separate the $|\pm 1\rangle$ states for a successful double-Lorentzian fitting. The circle is the measured result and the solid line represents the double-Lorentzian fitting result. **c.** The optical image of exfoliated MnBi$_2$Te$_4$ sample on the diamond substrate. The number of septuple layers (SLs) is labeled in the figure. Scale bar: 5 μm. **d.** The measured magnetic field $B_{NV}$ for different SLs of MnBi$_2$Te$_4$ sample. The "0-SLs" is collected at the surface of diamond as reference. The shadowed area represents the maximum error of the magnetic field $B_{NV}$.

- **S2.5.2. Nonlinear response in odd-layer MnBi$_2$Te$_4$ sample**

To further exclude the possibility that the nonlinear response in even-layer MnBi$_2$Te$_4$ are originated from residual magnetization, in this section, we investigate the nonlinear response in the 5SL-MnBi$_2$Te$_4$ device. All measurements were carried out in the absence of a magnetic field. Fig. S6 displays the linear and nonlinear longitudinal and transverse voltages in the 5SL-MnBi$_2$Te$_4$ as a function of applied current. In the linear regime, the longitudinal $V_x^\omega$ and transverse $V_y^\omega$ exhibit linear dependence on the applied current. In the nonlinear regime, the second-order longitudinal voltage $V_x^{2\omega}$ shows a weak quadratic dependence on the applied current, while the second-order transverse voltage $V_y^{2\omega}$ remains nearly zero. That is, only the longitudinal nonlinear response is observed in odd-layer MnBi$_2$Te$_4$, while the transverse response is absent. In contrast, there is a strong nonlinear response in both longitudinal and transverse directions in even-layer MnBi$_2$Te$_4$. Fig. S7 compares the nonlinear transverse response in 5SL and 4SL MnBi$_2$Te$_4$. Although the 5SL-MnBi$_2$Te$_4$ has a large net magnetization, its nonlinear transverse response is absent, which further supports the conclusion that the nonlinear response observed in even-layer MnBi$_2$Te$_4$ does not originate from residual magnetization.

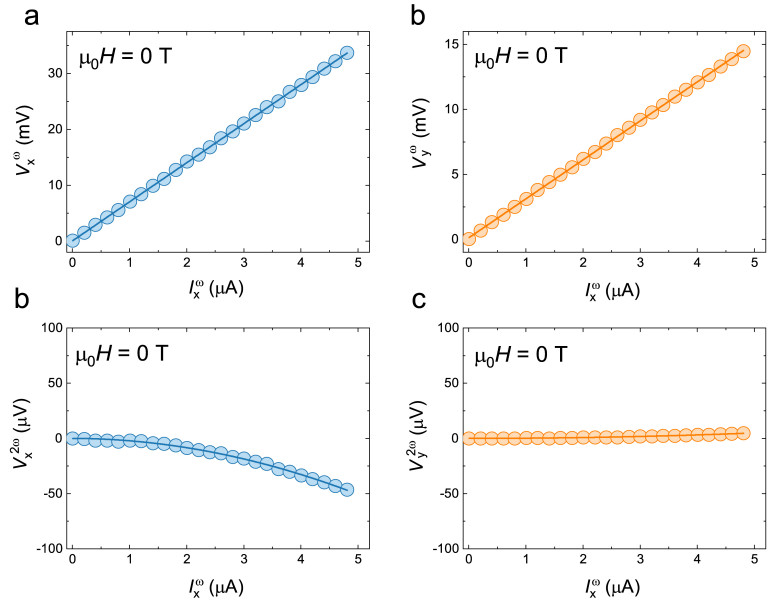

**Fig. S6 The linear and nonlinear response in 5SL-MnBi$_2$Te$_4$ a** and **b.** The linear longitudinal $V_x^\omega$ and transverse $V_y^\omega$ voltages as a function of applied current $I_x^\omega$. **c** and **d.** The second order nonlinear longitudinal $V_x^{2\omega}$ and transverse $V_x^{2\omega}$ voltages as a function of applied current $I_x^\omega$. All the measurement are carried out at $\mu_0 H = 0$ T and $T = 2$ K.

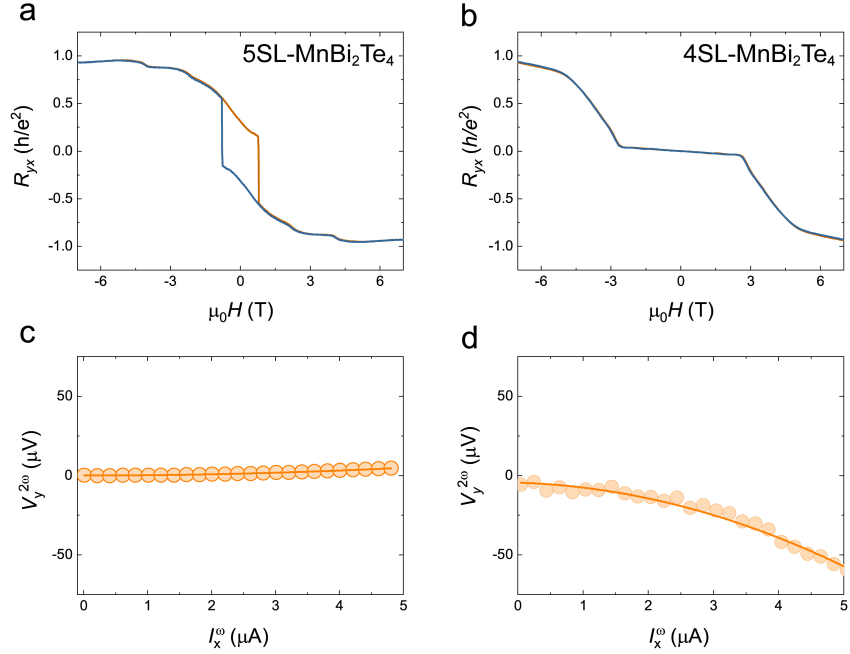

**Fig. S7 The comparison of the nonlinear Hall effect in the odd-layer and even-layer MnBi$_2$Te$_4$ a** and **b** The linear Hall resistance as a function of applied magnetic field for 5SL and 4SL MnBi$_2$Te$_4$ devices. **c** and **d**. The second order nonlinear Hall voltage as a function of applied current $I_x^\omega$ for 5SL and 4SL MnBi$_2$Te$_4$ devices.

Next, we investigate the observed nonlinear longitudinal response in the 5SL-MnBi$_2$Te$_4$ and demonstrate that it is fundamentally different from the nonlinear longitudinal response observed in the 4SL-MnBi$_2$Te$_4$, which is attributed to the quantum metric dipole. As shown in Fig. S8, we measured the nonlinear longitudinal resistance as a function of carrier density for 5SL and 4SL MnBi$_2$Te$_4$ at different edges. For the 5SL-MnBi$_2$Te$_4$, the nonlinear longitudinal resistance reverses its sign when we change the edge position (Fig. S8a). In the doped quantum anomalous Hall insulator (QAHI) (e.g., odd-layer MnBi$_2$Te$_4$), chiral edge state hybridizes strongly with trivial edge states, leading to asymmetric dispersion between opposite momenta, i.e., different magnitudes of Fermi velocities along opposite directions[5]. This velocity asymmetry coincides with the fact that both inversion symmetry and time-reversal symmetry are broken on the edge. Assuming a finite relaxation time, the direction-dependent mean free path arises from the direction-dependent Fermi velocity and eventually results in direction-dependent resistance. The nonreciprocal resistance in odd-layer MnBi$_2$Te$_4$ can be described based on a phenomenological model as $V_{xx} = iR_{xx} = iR_0 + \gamma R_0 i^2 (\hat{M} \times \hat{P}) \cdot \hat{i}$, where $R_0$ is the resistance that does not change with the current, $\gamma$ is the constant that characterizes the strength of the non-reciprocal charge transport effect, $\hat{M}$ is the magnetization direction of the $MnBi_2Te_4$, $\hat{P}$ is the edge charge dipole which is opposite on two edges, and $\hat{i}$ is the current direction. In stark contrast, the nonlinear longitudinal response in the even-layer $MnBi_2Te_4$ does not show any edge dependence, as shown in Fig. S8b. This result further confirms that the nonlinear response observed in even-layer $MnBi_2Te_4$ is not derived from residual magnetization.

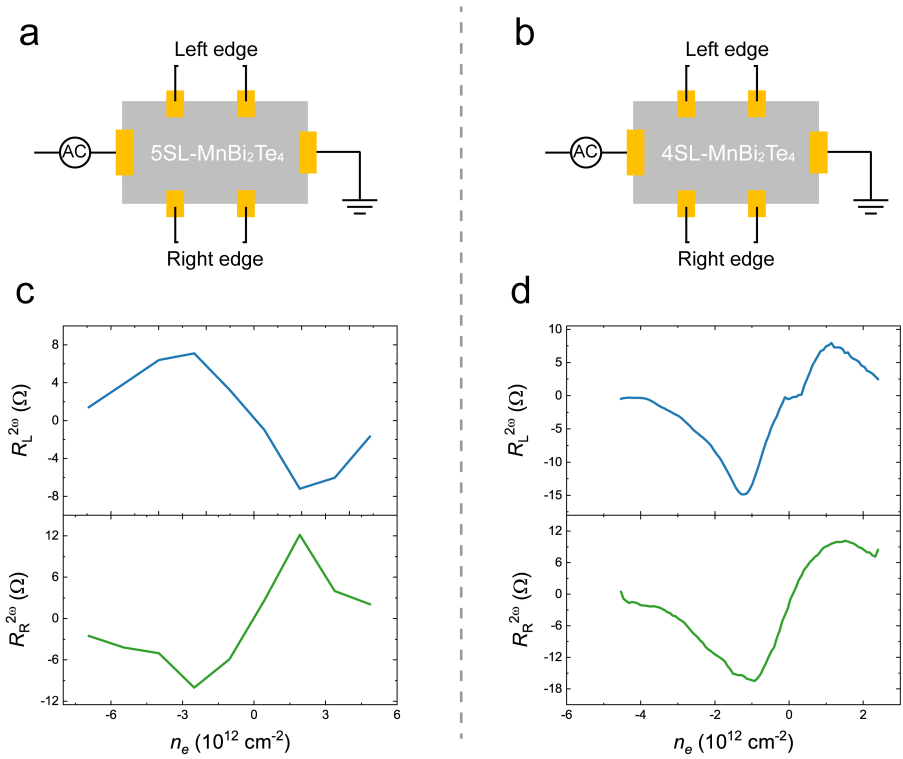

**Fig. S8 The comparison of the nonreciprocal longitudinal response in the odd-layer and even-layer $MnBi_2Te_4$ a** and **b.** The schematic illustration of the longitudinal response measurement in the odd-layer and even-layer $MnBi_2Te_4$. **c**. The carrier density dependent nonreciprocal longitudinal resistance for 5SL-$MnBi_2Te_4$, measured at the left and right edges. **d.** The carrier density dependent nonreciprocal longitudinal resistance for 4SL-$MnBi_2Te_4$, measured at the left and right edges.

In summary, we have carefully considered and excluded the possibility that the nonlinear response observed in even-layer $MnBi_2Te_4$ originated from residual magnetization.

## S3. Relationship between the magnetic states and the nonlinear response in even-layer MnBi$_2$Te$_4$

In the below, we will show the relationship between the magnetic state and the nonlinear response. We first present an analysis of the magnetic states in detail for the magnetic field dependent Hall resistance[6]. As depicted in Fig. S9a, at $\mu_0 H = 0$ T, the magnetic moment (↓↑↓↑) is fully compensated with the magnetic moment in the adjacent layers antiparallel to each other. Upon increasing the magnetic field to the first spin-flop field ($\mu_0 H_1 \approx 2$T), the magnetic moment configuration deviates from the equilibrium position, transitioning to the state of ↑↓↑↑. With a further increase in the magnetic field beyond the next spin-flop field ($\mu_0 H_1' \approx 3.5$T), the magnetization rises, and eventually, at $\mu_0 H_2 \approx 7$T, all the magnetic moments align in the direction of the external magnetic field with a configuration of ↑↑↑↑.

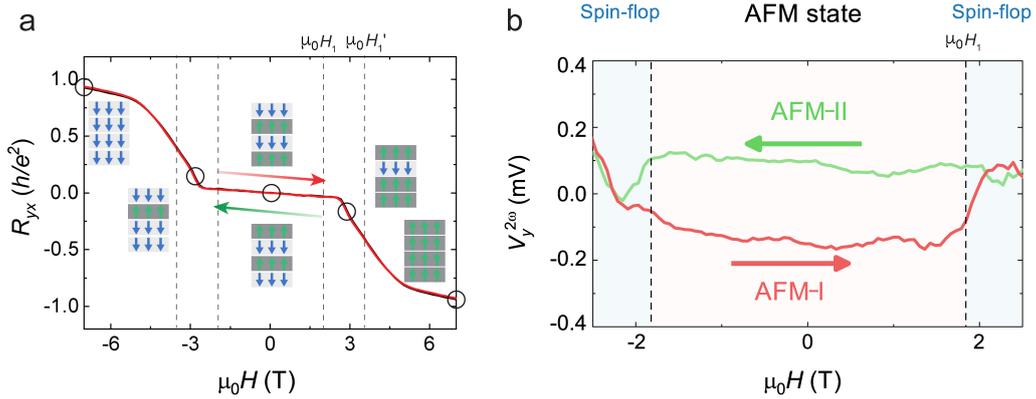

**Fig. S9 The relationship between the nonlinear response and the magnetic states in 4SL-MnBi$_2$Te$_4$. a.** The Hall resistance plotted against applied magnetic field for 4SL-MnBi$_2$Te$_4$ with magnetic states illustrated as cartoons at representative magnetic field values (black circles). Two spin-flop magnetic fields, $\mu_0 H_1$ and $\mu_0 H_1'$, are shown as dashed lines. **b.** The nonlinear transverse voltage as a function of magnetic field. The corresponding magnetic states are labeled in the figure. $\mu_0 H_1$ is the spin-flop magnetic field.

Here, we provide a more detailed explanation to demonstrate the relationship between the nonlinear response and the magnetic states. In Fig.S9b, the loop of the nonlinear response corresponds to the AFM states of even-layer MnBi$_2$Te$_4$. As the magnetic field increases to the spin-flop field $\mu_0 H_1$, the magnetic states of even-layer MnBi$_2$Te$_4$ enter the spin-flop state and the loop of the nonlinear response gradually vanishes. Because the spin-flop state exhibits randomly oriented spins in the plane, the nonlinear responses vanish because of averaging all directions.

Fig. S10 shows the magnetic field dependence of the nonlinear voltage $V_{yxx}^{2\omega}$ and $V_{xxx}^{2\omega}$. The difference between these values is only observed in the AFM states of MnBi$_2$Te$_4$. Additionally, the nonlinear response is nearly constant in the AFM region, indicating its robustness against the small magnetic field perturbations.

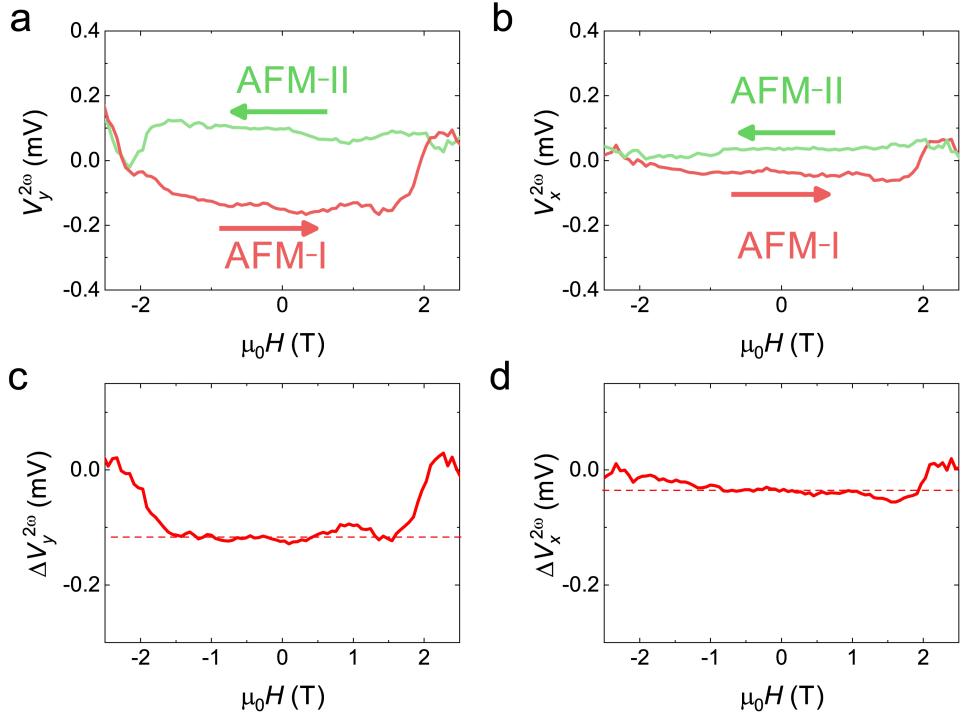

**Fig. S10 The magnetic field dependent nonlinear signal**. **a** and **b**. The nonlinear transverse $V_{yxx}^{2\omega}$ and longitudinal $V_{xxx}^{2\omega}$ voltages as a function of magnetic field. The nonlinear voltages are opposite for different AFM states. **c** and **d**. The difference of nonlinear transverse $\Delta V_{yxx}^{2\omega}$ and longitudinal $\Delta V_{xxx}^{2\omega}$ voltages between two AFM states as a function of magnetic field. Here, the $\Delta V^{2\omega} = (V^{2\omega}(AFM\ I) - V^{2\omega}(AFM\ II))/2$.

Under a high magnetic field, the spin moment of 4SL-MnBi$_2$Te$_4$ aligns in the same direction, and the quantum anomalous Hall effect could be observed in 4SL-MnBi$_2$Te$_4$. Fig S11 shows the nonlinear longitudinal and transverse voltage of 4SL-MnBi$_2$Te$_4$ when it reaches the quantized state under a large out-of-plane magnetic field. As depicted in Fig. S11 a and b, both the nonlinear longitudinal and transverse voltage vanishes. In the quantized state, the Dirac surface states of MnBi$_2$Te$_4$ become fully gapped. Since the quantum metric is closely related to the Fermi surface, the nonlinear longitudinal and transverse voltage disappears. Furthermore, from a symmetry perspective, when the magnetization of each layer in MnBi$_2$Te$_4$ aligns in the same direction, even-layer MnBi$_2$Te$_4$ breaks time reversal symmetry ($T$) but maintains inversion symmetry ($P$). As the net quantum metric dipole requires the breaking of both time reversal symmetry ($T$) and inversion symmetry ($P$), the nonlinear response in even-layer MnBi$_2$Te$_4$ vanishes under the high magnetic field.

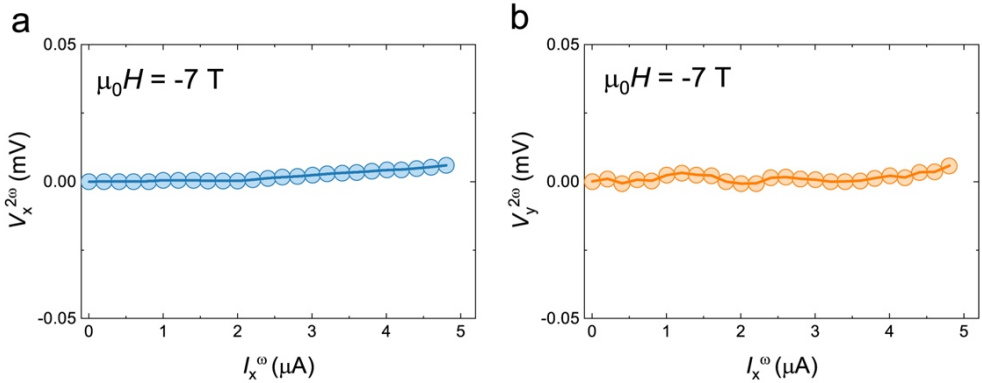

**Fig. S11 Nonlinear Response under quantized states in 4SL-MnBi$_2$Te$_4$ a.** The second order nonlinear longitudinal voltage as a function of current. **b.** The second order nonlinear transverse voltage as a function of current. All the measurement are carried out at the magnetic field $\mu_0 H = -7$ T and temperature $T = 1.6$ K.

**S4. The nonreciprocal longitudinal response in 4SL-MnBi$_2$Te$_4$**

In this section, we will give a detailed explanation on the definition of nonreciprocal longitudinal response in 4SL-MnBi$_2$Te$_4$. We will also provide additional DC measurement result to present the nonreciprocal longitudinal response in 4SL-MnBi$_2$Te$_4$. Moreover, we will evaluate the nonreciprocal coefficient in 4SL-MnBi$_2$Te$_4$.

## ❖ S4.1 Definition of the nonreciprocal longitudinal response in 4SL-MnBi$_2$Te$_4$

The " nonreciprocal longitudinal response" in 4SL-MnBi$_2$Te$_4$ represents that the longitudinal resistance is asymmetric when reversing the current direction, which can be described as $R(I) \neq R(-I)$, where $R$ is the resistance and $I$ is the current direction, as shown in Fig. S12. This behavior is similar to a diode, whose conductance is asymmetric with different injecting current directions. The term "non-reciprocal" means that the longitudinal resistance violates Onsager's relation. In general, Onsager's relations predict that the transport coefficients (e.g., thermal conductivity, electrical conductivity, and diffusion) in a linear, near-equilibrium system are symmetric. This means that the resistance experienced when applying a current in one direction is equal to the resistance experienced when applying the current in the opposite direction. Namely, it can be described as $R(I) \neq R(-I)$ (Fig. S12). Obviously, the asymmetric longitudinal resistance ($R(I) \neq R(-I)$) violates Onsager's relation and is defined as "non-reciprocal".

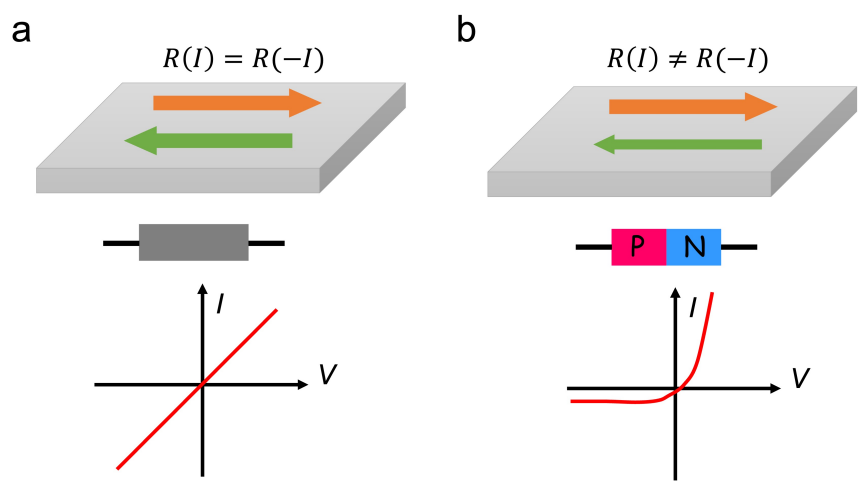

**Fig. S12 Schematic illustration of the reciprocal and nonreciprocal resistance a.** Schematic representation of reciprocal resistance ($R(I) = R(-I)$) with the corresponding *I-V* relationship displayed in the lower panel, demonstrating a linear dependence. **b.** Schematic representation of nonreciprocal resistance ($R(I) \neq R(-I)$) with the corresponding *I-V* relationship displayed in the lower panel, illustrating deviation from linear dependence.

In the even-layer MnBi$_2$Te$_4$, due to the quantum metric dipole, a nonlinear longitudinal response which is quadric to the applied electric field will emerge, described as $J_x^{2\omega} = \sigma_{xxx}^{2\omega} E_x^2$. According to this, we can rewrite the *I-V* characteristic of even-layer MnBi$_2$Te$_4$ as[7]:

$$V_x/l = iR_0 + \gamma R_0 i^2 \qquad (Eq.\ S1)$$

, where $V_x$ is the voltage drop of the contact measured at the longitudinal direction, $l$ is the length between the contacts, $i$ is the applied current density, $R_0$ is the linear resistance, $\gamma$ is the coefficient evaluating the magnitude of the nonreciprocity. Therefore, based on Eq. S1, we can get that the longitudinal resistance in even-layer MnBi$_2$Te$_4$ is asymmetric when reversing the current direction and is then defined as "nonreciprocal longitudinal response". The difference in longitudinal resistance when reversing the current direction can be described as $\Delta R_{xx} = R(+i) - R(-i) = 2\gamma R_0 i^{DC}$. In the AC measurement, by applying an AC current $i = \sqrt{2} i^{RMS} \sin(\omega t)$, the longitudinal voltage drop can be written as

$$V_x/l = \sqrt{2} \sin(\omega t) i^{RMS} R_0 + \gamma R_0 (i^{RMS})^2 (1 + \sin(2\omega t - \frac{\pi}{2})) \qquad (Eq.S2)$$

Here, we can define the "nonreciprocal longitudinal resistance" $R_{xx}^{2\omega}$ as $R_{xx}^{2\omega} = \frac{V_x^{2\omega}}{i^{RMS}}$. Therefore, we have $\Delta R_{xx} = 2\sqrt{2} R_{xx}^{2\omega}$ and establish the connection between AC and DC measurements. Nonetheless, the AC method will demonstrate a better signal-to-noise ratio with a small current, which we employed in our current experiment.

### ❖ S4.2 DC measurement of the diode-like nonreciprocal longitudinal response in 4SL-MnBi$_2$Te$_4$

As we stated before, the nonreciprocal longitudinal response in 4SL-MnBi$_2$Te$_4$ can be directly observed by DC measurement, where a nonlinear DC *I-V* characteristic caused by the nonlinearity brought by the quantum metric dipole should be observed. Moreover, since the quantum metric dipole-induced nonlinearity is antisymmetric under time reversal operation, we expect that the nonlinearity of the DC *I-V* characteristic is antisymmetric with opposite AFM states. In this section, we will present the DC measurement results for the 4SL-MnBi$_2$Te$_4$ device. Fig. S13a shows the four-wire DC *I-V* measurement results for the 4SL-MnBi$_2$Te$_4$ devices with opposite AFM I and AFM II states. It can be seen that the *I-V* relationship deviates from the linear relationship. Moreover, the nonlinearity is opposite for AFM I and AFM II states. To make it clearer, we extracted the difference in resistance $\Delta R_{xx} = R(+i) - R(-i)$, which is shown in Fig. S13b. It can be seen that the $\Delta R_{xx}$ shows a linear dependence on the current $I^{DC}$. Furthermore, the $\Delta R_{xx}$ has a similar value but shows an opposite sign for the AFM I and AFM II states, in good accordance with our expectation.

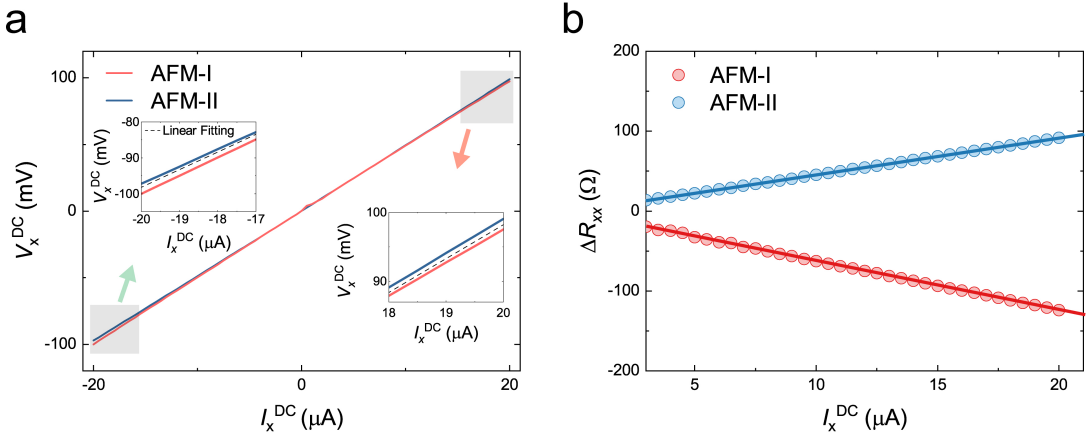

**Fig. S13 DC measurement of the 4SL-MnBi$_2$Te$_4$ device a.** The four wire DC *I-V* measurements of the 4SL-MnBi$_2$Te$_4$ device for AFM I and AFM II states. A clear nonlinearity of the *I-V* curve is observed. **b.** The difference in the longitudinal resistance $\Delta R_{xx}$ as a function of current.

❖ **S4.3 Evaluation of the nonreciprocal coefficient in 4SL-MnBi$_2$Te$_4$**

From Eq.S2, we obtain $\gamma = \sqrt{2} R_{xx}^{2\omega}/(R_0 \cdot i_0^{RMS})$, which can evaluate the magnitude of the nonreciprocity. The coefficient $\gamma$ as a function of carrier density is shown in Fig. S14. As depicted in Fig. S14, the nonreciprocal coefficient $\gamma$ of 4SL-MnBi$_2$Te$_4$ can reach a maximum value of $\gamma \approx 7 \times 10^{-11} m^2 A^{-1}$. This value is at least two or three orders larger than traditional heavy metal (HM)/ferromagnetic metal (FM) heterostructure systems, such as Pt/Co or Pt/Py[8,9]. Moreover, it is also comparable to some 2D materials with broken inversion symmetry, like BiTeBr and WTe$_2$[10,11].

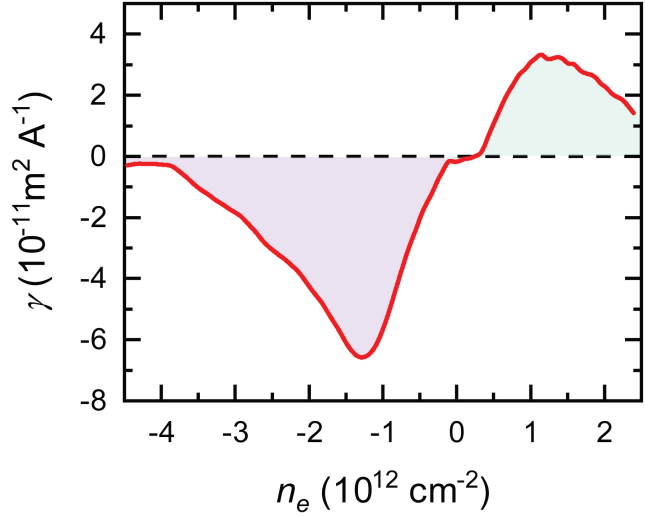

**Fig. S14 The nonreciprocal coefficient ($\gamma$) as a function of charge carrier density $n_e$ of 4SL-MnBi$_2$Te$_4$.**

## S5. Scaling relationship between the $|\frac{\sigma_{xx}}{n_e}|$ and temperature in 4SL-MnBi$_2$Te$_4$

To better support our claim that the conductivity $\sigma_{xx}$ is only relevant to the scattering time $\tau$ in the temperature range of 1.6 to 10 K while the charge carrier density $n_e$ remains unchanged in this temperature range, we plot the $|\frac{\sigma_{xx}}{n_e}|$ as a function of temperature, as shown in Fig. S15. In the temperature range of 1.6 K to 10 K, the $|\frac{\sigma_{xx}}{n_e}|$ exhibits a similar temperature dependence to that of the conductivity, supporting our claim that the scattering time "$\tau$" is the main component that dominates the temperature dependence of the conductivity $\sigma_{xx}$ in this temperature range.

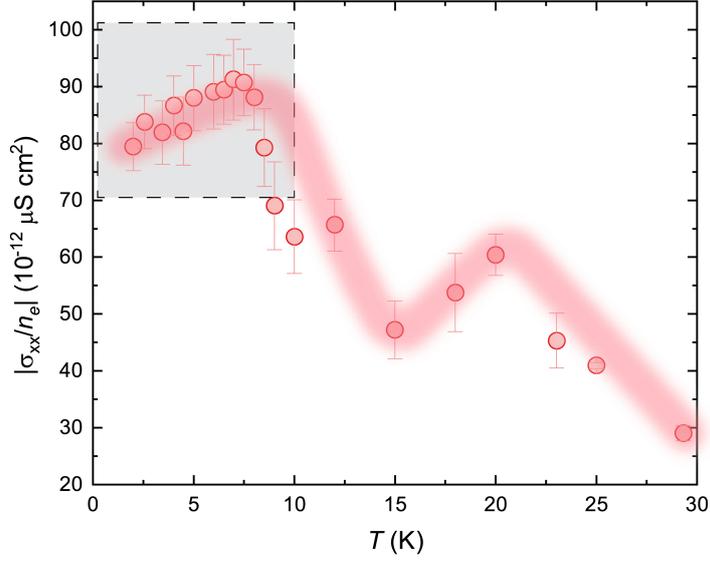

**Fig. S15** The $\left|\frac{\sigma_{xx}}{n_e}\right|$ as a function of temperature for the 4SL-MnBi$_2$Te$_4$ device.

## S6. Additional data from a 6SL-MnBi$_2$Te$_4$ sample

We carried out the nonlinear transport measurement on an additional 6SL-MnBi$_2$Te$_4$ device. Fig. S16a and b shows device picture of the 6SL-MnBi$_2$Te$_4$. For the 6SL-MnBi$_2$Te$_4$, we also observed large second-harmonic transverse ($V_y^{2\omega}$) and longitudinal ($V_x^{2\omega}$) voltage at zero field. Additionally, as shown in Fig. S16 c and d, similar to the 4SL-MnBi$_2$Te$_4$ device, we also observed the sign reversal of the second-harmonic transverse ($V_y^{2\omega}$) and longitudinal ($V_x^{2\omega}$) voltage in 6SL-MnBi$_2$Te$_4$ when preparing the different AFM states (i.e. AFM-I and AFM-II). We then measured the temperature dependence of the nonlinear transverse voltage $V_y^{2\omega}$ and longitudinal voltage $V_x^{2\omega}$ of the 6SL-MnBi$_2$Te$_4$ device, as shown in Fig. S17. Both $V_y^{2\omega}$ and $V_x^{2\omega}$ decreases with temperature and vanished when above the Neél temperature of MnBi$_2$Te$_4$, consistent with the result from 4SL-MnBi$_2$Te$_4$.

Next, we investigate the scaling relationship between $\sigma_{yxx}^{2\omega}$ and $\sigma_{xxx}^{2\omega}$ and conductivity $\sigma_{xx}^{\omega}$ in 6SL-MnBi$_2$Te$_4$ by varying the temperature, as shown in Fig. S18. By fitting the scaling plot with the formula $\sigma^{2\omega} = \eta_2(\sigma_{xx}^{\omega})^2 + \eta_0$, we found that the most prominent contribution for $\sigma_{yxx}^{2\omega}$ and $\sigma_{xxx}^{2\omega}$ is the $\eta_0$ term, which is independent of the scattering time and originates from the normalized quantum metric dipole. This result is also consistent with the that in 4SL-$MnBi_2Te_4$.

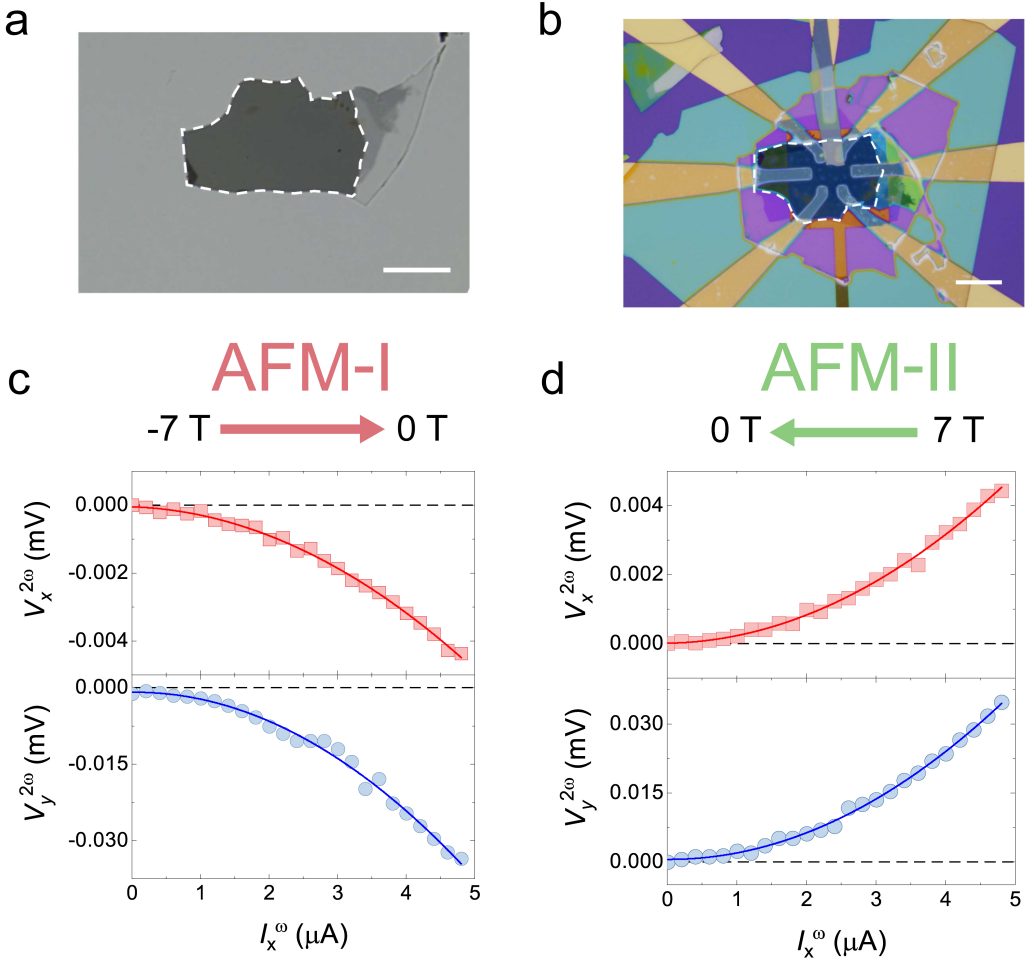

**Fig. S16. The spin order related nonlinear signals in 6SL-$MnBi_2Te_4$. a.** The transmitted optical image of exfoliated 6SL-$MnBi_2Te_4$ thin flakes on $Al_2O_3$ thin film. The $MnBi_2Te_4$/$Al_2O_3$ is supported by PDMS. Scale bar: 20 μm. **b.** The optical image of the device from the sample shown in **a**. Cr/Au electrodes are deposited on 6SL-$MnBi_2Te_4$ with stencil mask and then covered by hBN/graphite as the top gate. The white dashed line represents the shape of 6SL-$MnBi_2Te_4$. Scale bar: 20 μm. **c and d.** The nonlinear longitudinal $V_x^\omega$ and transverse $V_y^\omega$ voltage as a function of current $I_x^\omega$ for AFM-I and AFM-II states, respectively. The solid lines are quadratic fittings for the data.

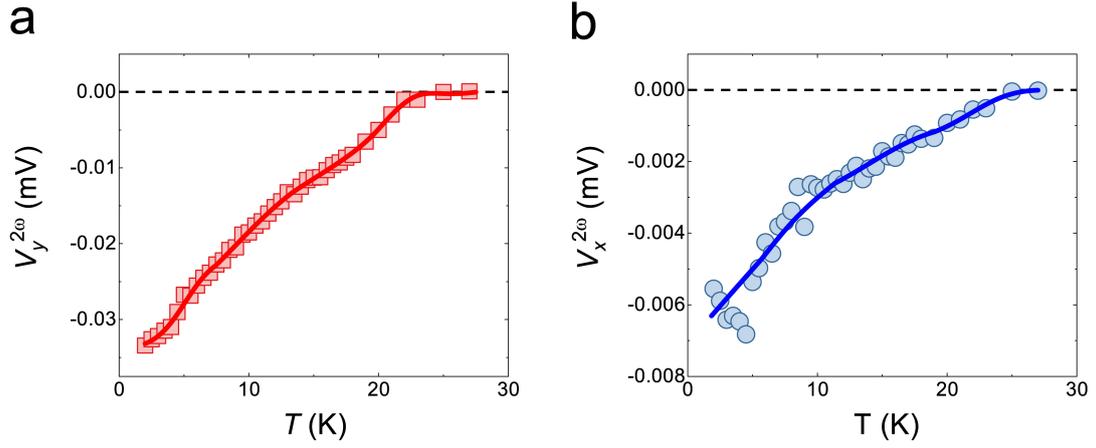

**Fig. S17.** The nonlinear transverse voltage $V_y^{2\omega}$ and longitudinal voltage $V_x^{2\omega}$ of 6SL-MnBi$_2$Te$_4$ device as a function of temperature for AFM-I (corresponding to sweeping the magnetic field from -7 T to zero) state.

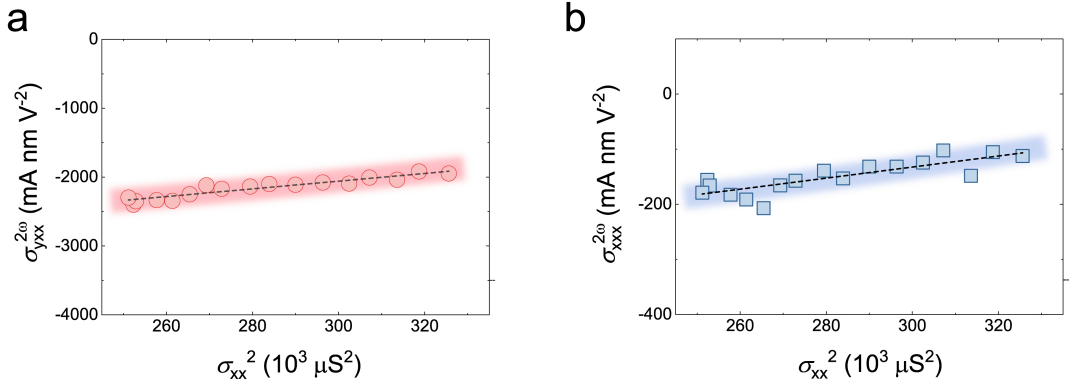

**Fig. S18.** The scaling relationship between the nonlinear conductivity $\sigma_{yxx}^{2\omega}$ ($\sigma_{xxx}^{2\omega}$) and the linear longitudinal conductivity $\sigma_{xx}$ for 6SL-MnBi$_2$Te$_4$. Dashed line is the fitting of the data with formula $\sigma^{2\omega} = \eta_2 (\sigma_{xx}^{\omega})^2 + \eta_0$.

Finally, we investigated the charge carrier density dependence on the nonlinear conductivity $\sigma_{yxx}^{2\omega}$ and $\sigma_{xxx}^{2\omega}$ in the 6SL-MnBi$_2$Te$_4$ device, as shown in Fig. S19. When the Fermi level is tuned into the charge neutrality gap, both $\sigma_{yxx}^{2\omega}$ and $\sigma_{xxx}^{2\omega}$ vanishes. Moreover, $\sigma_{yxx}^{2\omega}$ and $\sigma_{xxx}^{2\omega}$ shows sign change when tuning the carrier density from hole-doped regime to electron-doped regime. These observations are also consistent with that in 4SL-MnBi$_2$Te$_4$.

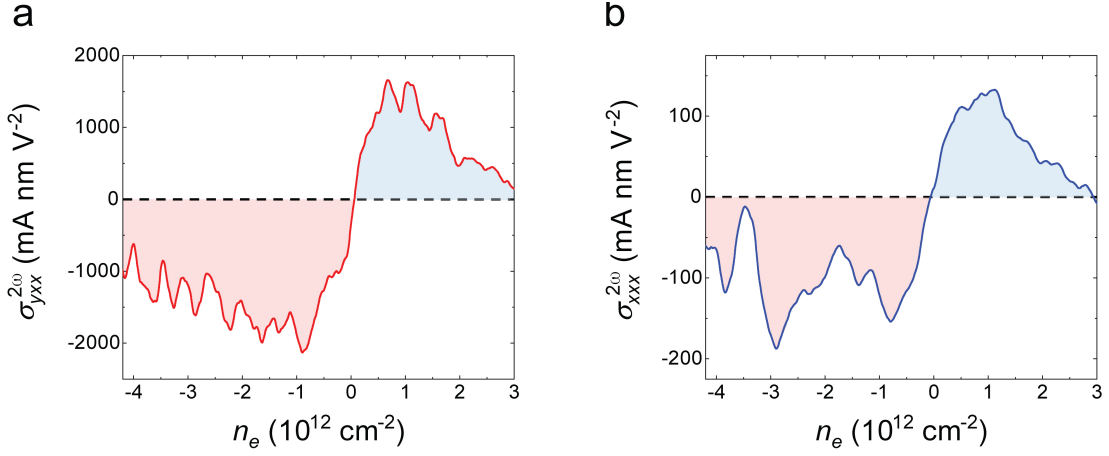

**Fig. S19. The measured nonlinear transverse conductivity $\sigma_{yxx}^{2\omega}$ and longitudinal conductivity $\sigma_{xxx}^{2\omega}$ as the function of charge carrier density $n_e$.**

**S7. The influence of disorder and other mechanism on the nonlinear response in MnBi$_2$Te$_4$**

In this section, we elaborate further on the theoretical calculation about the influence of disorder on the quantum metric dipole induced nonlinear response in MnBi$_2$Te$_4$. We will first discuss the correlation between the intrinsic band gap and nonlinear conductivity. Then, we will discuss the contribution of scattering to the nonlinear conductivity. Finally, we will discuss all the possible mechanism that can contribute to the nonlinear response and explain why they can be excluded from the symmetry aspect as well as the scattering time $\tau$ dependence.

❖ **S7.1 Variation of the nonlinear conductivity**

In this section, we focus one contribution to the magnitude of the nonlinear conductivity – the intrinsic gap of MnBi$_2$Te$_4$, aiming to explain the discrepancy between the theoretically calculated value of the nonlinear conductivity and the experimentally obtained values. Although the recent experiments have all reported the absence of any magnetic gap on the surface of MnBi$_2$Te$_4$, *ab initio* calculations, however, consistently show a robust magnetic gap; in 4SL MnBi$_2$Te$_4$ we obtain a value of $E_g = 60\ meV$ [12-14]. A recent study has demonstrated that co-antisite defects in Mn-Bi sites – while maintaining the magnetic order – significantly suppress the surface gap[15]. As such defects are known to be abundant in MnBi$_2$Te$_4$, it is plausible that such a gap suppression occurs in our samples[16,17]. The quantum metric may be enhanced by the shrinking magnetic gap and modified surface band structure in experiments.

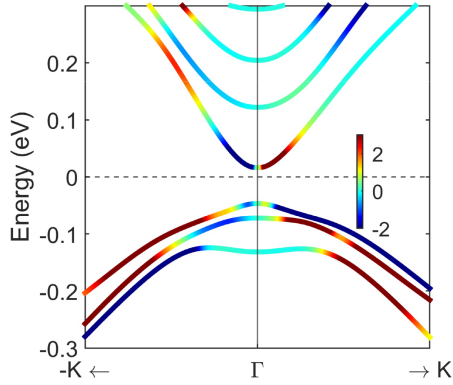

**Fig. S20 Band structure and nonlinear conductivity**. Band structure and projected nonlinear conductivity for the pristine phase. Color bar scale is in mA nm V$^{-2}$.

❖ **S7.2. Contribution of scattering to the nonlinear conductivity**

Impurities, both magnetic and non-magnetic, have been recognized to influence nonlinear conductivity, beyond the intrinsic sources discussed in our work. In this section, we review scattering mechanisms and exclude their contribution to the intrinsic conductivity (*i.e.* nonlinear conductivity scaling as $\tau^0$).

We first note that that rotational symmetry can suppress certain forms of skew scattering. Prior research has suggested that skew scattering, which is typically important in time-reversal symmetry breaking systems, may generate a sizeable nonlinear conductivity[18]. However, it relies on breaking all rotational symmetries in a 2D system. Consequently, it is negligible in MnBi$_2$Te$_4$ with well-preserved $C_{3z}$ symmetry. Intriguingly, under certain circumstances, the disorder scattering-induced nonlinear Hall effect can persist despite the presence of the $C_{3z}$ symmetry[19]. Nevertheless, it is important to highlight that the nonlinear conductivity from this origin strongly depends on the effective scattering rate $\tau$ [19]. In particular, skew-scattering contributes at order of $\tau^3$ and $\tau^2$, whereas side-jump scales with $\tau^2$ and $\tau$. The scaling analysis in our main text ($\sigma^{2\omega} = \eta_2(\sigma_{xx}^{\omega})^2 + \eta_0$) rules out such contributions, and the intrinsic conductivity at order $\tau^0$ is unaffected by these disorder terms.

Additionally, based on the Boltzmann formalism, it has been proposed that the contributions to the asymmetric scattering rate at nonlinear have the general form $\tau^{n-1}E^n$ [20]. The nonlinear conductivity is obtained for $n > 1$, where the lowest order disorder contribution begins at $\tau^1$. We therefore conclude that existing impurity scattering mechanisms cannot explain the dominant τ-independent nonlinear conductivity.

**S7.3. All the possible mechanism to the nonlinear response**

To support our argument further, we provide all the mechanisms that can contribute to the nonlinear transport and explain why their contribution is excluded in even-layer $MnBi_2Te_4$.

Table I. Summary of the scattering time $\tau$ dependence and the symmetry analysis for different mechanism (" × " means forbidden, "✓" means allowed)

| Mechanism | $\tau$-dependence | Symmetry | | |
|---|---|---|---|---|
| | | P | T | PT |
| Berry curvature dipole[21] | $\tau^1$ | × | ✓ | × |
| Skew scattering[19,20] | $\tau^3$ or $\tau^2$ | × | ✓ | × |
| Side jump[19] | $\tau^2$ or $\tau^1$ | × | ✓ | × |
| Anomalous skew scattering[18] | $\tau^2$ | × | × | ✓ |
| Second order Drude term[10] | $\tau^2$ | × | × | ✓ |
| Quantum metric dipole | $\tau^0$ | × | × | ✓ |

Given the symmetry of even-layer $MnBi_2Te_4$, which breaks the inversion symmetry (P) and time-reversal symmetry (*T*) but preserves the combined *PT* symmetry, we can exclude the contribution from Berry curvature dipole, skew scattering, side jump. The anomalous skew scattering contribution is suppressed by rotational symmetry of even-layer MnBi$_2$Te$_4$. More importantly, based on the scaling relationship shown in Fig. 3d, the $\tau^0$- dependent quantum metric dipole contribution dominate the nonlinear conductivity, while the $\tau^2$- dependent Drude term is negligible. Therefore, we conclude that the quantum metric dipole is the dominant contributor to the observed nonlinear response in the even-layer MnBi$_2$Te$_4$.

**References:**

1. Rondin, L. *et al.* Magnetometry with nitrogen-vacancy defects in diamond. *Rep. Prog. Phys.* **77**, 056503 (2014).

2. Gross, I. *et al.* Real-space imaging of non-collinear antiferromagnetic order with a single-spin magnetometer. *Nature* **549**, 252-256 (2017).

3. Thiel, L. *et al.* Probing magnetism in 2D materials at the nanoscale with single-spin microscopy. *Science* **364**, 973-976 (2019).

4. McLaughlin, N. J. *et al.* Quantum Imaging of Magnetic Phase Transitions and Spin Fluctuations in Intrinsic Magnetic Topological Nanoflakes. *Nano Lett.* **22**, 5810-5817 (2022).

5. Zhang, Z. *et al.* Controlled large non-reciprocal charge transport in an intrinsic magnetic topological insulator $MnBi_2Te_4$. *Nat. Commun.* **13**, 6191 (2022).

6. Yang, S. *et al.* Odd-Even Layer-Number Effect and Layer-Dependent Magnetic Phase Diagrams in $MnBi_2Te_4$. *Phys. Rev. X* **11**, 011003 (2021).

7. Tokura, Y. & Nagaosa, N. Nonreciprocal responses from non-centrosymmetric quantum materials. *Nat. Commun.* **9**, 3740 (2018).

8. Avci, C. O., Mendil, J., Beach, G. S. D. & Gambardella, P. Origins of the Unidirectional Spin Hall Magnetoresistance in Metallic Bilayers. *Physical Review Letters* **121**, 087207 (2018).

9. Avci, C. O. *et al.* Unidirectional spin Hall magnetoresistance in ferromagnet/normal metal bilayers. *Nature Physics* **11**, 570-575 (2015).

10. Ideue, T. *et al.* Bulk rectification effect in a polar semiconductor. *Nature Physics* **13**, 578-583 (2017).

11. He, P. *et al.* Nonlinear magnetotransport shaped by Fermi surface topology and convexity. *Nature Communications* **10**, 1290 (2019).

12. Hao, Y.-J. *et al.* Gapless Surface Dirac Cone in Antiferromagnetic Topological Insulator $MnBi_2Te_4$. *Phys. Rev. X* **9**, 041038 (2019).

13. Li, H. *et al.* Dirac Surface States in Intrinsic Magnetic Topological Insulators $EuSn_2As_2$ and $MnBi_{2n}Te_{3n+1}$. *Phys. Rev. X* **9**, 041039 (2019).

14. Hu, Y. *et al.* Universal gapless Dirac cone and tunable topological states in $(MnBi_2Te_4)_m(Bi_2Te_3)_n$ heterostructures. *Phys. Rev. B* **101**, 161113 (2020).

15. Tan, H. & Yan, B. Distinct Magnetic Gaps between Antiferromagnetic and Ferromagnetic Orders Driven by Surface Defects in the Topological Magnet $MnBi_2Te_4$. *Phys. Rev. Lett.* **130**, 126702 (2023).

16. Liu, Y. *et al.* Site Mixing for Engineering Magnetic Topological Insulators. *Phys. Rev. X* **11**, 021033 (2021).

17. Garnica, M. *et al.* Native point defects and their implications for the Dirac point gap at $MnBi_2Te_4$(0001). *npj Quantum Mater.* **7**, 7 (2022).

18. Ma, D., Arora, A., Vignale, G. & Song, J. C. Anomalous skew-scattering nonlinear Hall effect *PT*-symmetric antiferromagnets. *arXiv:2210.14932* (2022).

19. Du, Z. Z., Wang, C. M., Li, S., Lu, H.-Z. & Xie, X. C. Disorder-induced nonlinear Hall effect with time-reversal symmetry. *Nat. Commun.* **10**, 3047 (2019).

20. Isobe, H., Xu, S.-Y. & Fu, L. High-frequency rectification via chiral Bloch electrons. *Sci. Adv.* **6**, eaay2497 (2020).
50